\documentclass[sigconf]{acmart}
\pdfoutput=1

\usepackage{booktabs}
\usepackage{graphicx}
\usepackage{balance}
\usepackage{xspace}
\usepackage{caption}
\usepackage[T1]{fontenc}
\usepackage{wrapfig}
\usepackage[tableposition=top,font={scriptsize},figurename=Fig., skip=7pt]{caption}

\def\eg/{e.g.,}
\def\etc/{etc.}
\def\etal/{et al.}
\def\ie/{i.e.,}

\newcommand{\Identifiers}{Publisher-specific\,IDs\xspace}

\newcommand{\identifiers}{publisher-specific\,IDs\xspace}
\newcommand{\publisherID}{\texttt{Publisher\,ID}\xspace}
\newcommand{\containerID}{\texttt{Container\,ID}\xspace}
\newcommand{\trackingID}{\texttt{Tracking\,ID}\xspace}
\newcommand{\measurementID}{\texttt{Measurement\,ID}\xspace}
\newcommand{\analyticsID}{\texttt{Analytics\,ID}\xspace}
\newcommand{\publisherIDs}{\texttt{Publisher\,IDs}\xspace}
\newcommand{\containerIDs}{\texttt{Container\,IDs}\xspace}
\newcommand{\trackingIDs}{\texttt{Tracking\,IDs}\xspace}
\newcommand{\measurementIDs}{\texttt{Measurement\,IDs}\xspace}
\newcommand{\analyticsIDs}{\texttt{Analytics\,IDs}\xspace}

\newcommand{\metagraph}{metagraph\xspace}
\newcommand{\Metagraph}{Metagraph\xspace}

\newcommand{\adstxt}{\texttt{ads.txt}\xspace}

\newcommand{\trancoCrawl}{1MT crawl\xspace}

\newcommand{\historicalSites}{8M websites\xspace}

\begin{document}
\pagestyle{plain}

\copyrightyear{2022}
\acmYear{2022}
\setcopyright{acmcopyright}
\acmConference[WWW '22]{Proceedings of the ACM Web Conference 2022}{April 25--29, 2022}{Virtual Event, Lyon, France}
\acmBooktitle{Proceedings of the ACM Web Conference 2022 (WWW '22), April 25--29, 2022, Virtual Event, Lyon, France}
\acmPrice{15.00}
\acmDOI{10.1145/3485447.3512124}
\acmISBN{978-1-4503-9096-5/22/04}

\title{Leveraging Google's Publisher-specific IDs to\\ Detect Website Administration}

\author{Emmanouil Papadogiannakis}
\affiliation{
	\institution{FORTH/University of Crete}
	\country{Greece}
}
\author{Panagiotis Papadopoulos}
\affiliation{
	\institution{Telefonica Research}
	\country{Spain}
}
\author{Evangelos P. Markatos}
\affiliation{
	\institution{FORTH/University of Crete}
	\country{Greece}
}
\author{Nicolas Kourtellis}
\affiliation{
	\institution{Telefonica Research}
	\country{Spain}
}



\begin{abstract}
Digital advertising is the most popular way for content monetization on the Internet.
Publishers spawn new websites, and older ones change hands with the sole purpose of monetizing user traffic.
In this ever-evolving ecosystem, it is challenging to effectively answer questions such as: Which entities monetize what websites?
What categories of websites does an average entity typically monetize on and how diverse are these websites?
How has this website administration ecosystem changed across time?

In this paper, we propose a novel, graph-based methodology to detect administration of websites on the Web, by exploiting the ad-related \identifiers.
We apply our methodology across the top 1 million websites and study the characteristics of the created graphs of website administration.
Our findings show that approximately 90\% of the websites are associated each with a single publisher, and that small publishers tend to manage less popular websites.
We perform a historical analysis of up to 8 million websites, and find a new, constantly rising number of (intermediary) publishers that control and monetize traffic from hundreds of websites, seeking a share of the ad-market pie.
We also observe that over time, websites tend to move from big to smaller administrators.
\end{abstract}

\maketitle

\section{Introduction}
\label{sec:introduction}

Digital advertising keeps the content we consume on the Web free of charge, being an important stream of revenue for web publishers~\cite{adSpending}. 
Even during 2020, with all the adverse economic impacts of the COVID-19 pandemic, there was a reported 12.2\% increase in ad-revenues~\cite{iabAdRevenue-covid-2020}, and \$100s of billions in annual spending worldwide (\$455B in 2021~\cite{adSpending}).
However, it is inherently difficult to assess the effectiveness of digital ad spending, due to the overly complex, layered ecosystem of digital marketing, with thousands of intermediaries brokering ads and ad-slots between sellers and buyers.
Some even consider this market overvalued and possibly due to correction with various adverse effects~\cite{hwang2020adtimebomb}.

In an attempt to increase ad profits, advertisers, intermediaries and publishers resort to analytics and other web tracking services for better measuring of user audiences and their engagement with webpages.
But the increasing complexity of this ecosystem makes it hard to answer questions such as:
Who are the entities that control and monetize websites, and which websites?
How many websites does the average such entity control? 
Are they from the same category, or diverse in nature? 
What are the characteristics of these website administrators, and how have these changed over time?

In the last decade, journalists and academic researchers have grappled with such questions.
In fact, they made several efforts to (i) provide more transparency to the ecosystem of web content monetization and administration~\cite{buzzfeed}, (ii) raise awareness of its impact to user's privacy due to online tracking and possibilities of de-anonymization~\cite{wired, matic2015caronte, starov2018betrayed,  bashir2019longitudinal, yoon2019doppelgangers}, (iii) shed light on how this ecosystem drives misinformation and fake news~\cite{globalVoices,buzzfeed,verafiles}. 
For example, L.~Alexander~\cite{globalVoices} used Google analytic IDs to find evidence of a pro-Kremlin concerted web campaign, executed among different websites owned by the same entity.
C.~Silverman \etal/~\cite{buzzfeed} looked into Google-related IDs and found websites being operated by the same entities, which promoted fake news content and delivered polarizing ads during the USA 2016 presidential election. 
Furthermore, C.I.~Samson~\cite{verafiles} discussed the issue of fake news spreading within the context of the Philippines 2016 presidential election.
Such reports demonstrate the urgent need for more transparency in the issue of website administration.
In addition, academic works~\cite{matic2015caronte, starov2018betrayed,  bashir2019longitudinal, yoon2019doppelgangers} have looked at the problem from the point of view of user tracking or de-anonymization using such Google-related IDs to detect malicious websites and their administrators. However, to date, there has been no systematic study to reveal, at scale, the way websites are monetized and by which entities.

In this work, we try to shed light on website administration and propose a novel, graph-based methodology to detect entities that are in charge of websites. 
To that extent, we exploit the ad-related, \identifiers that publishers embed in their websites in order to use third-party services. 
We (i) apply our methodology across the top 1 million websites of Tranco list, to detect groups of websites monetized by the same entity, (ii) study the characteristics of the generated website administration graphs, (iii) find \emph{intermediary} publishers that manage and monetize traffic from hundreds or even thousands of websites.
We perform a 2-year historical analysis of up to top 8 million websites and explore how the small, medium and large publishers have evolved over time.

In summary, the contributions of this work are:
\begin{itemize}
    \item We propose a novel methodology for detecting website administration and co-ownership based on \identifiers, with applicability in different use cases.
    \item We conduct the first, to our knowledge, large-scale systematic study of such \identifiers embedded in up to \historicalSites. We make our implementation~\cite{openSourceCode} along with our results~\cite{openSourceData} publicly available to support further research on this topic. 
    \item Our findings show that approximately 90\% of the websites are associated with a single publisher and that small publishers tend to manage less popular websites. 
    We also conclude that there is preferential administration with an inclination towards ``News and Media'' websites. Finally, we show that over time, websites tend to move away from big administrators into smaller ones.
\end{itemize}

\section{\Identifiers}
\label{sec:background}

\subsection{Google AdSense}
\label{sec:googleAdSense}

\emph{AdSense} is a service for publishers to generate revenue by displaying ads in their websites. 
For ads to be displayed, publishers need to insert the AdSense code snippet in their website, which includes a \publisherID: a unique identifier for an AdSense account that follows the format \texttt{pub-XXXXXXXXXXXXXXX}. 
The owner of the account is allowed to share the account with employees, or even business partners, however, the account holder is always one, and different AdSense accounts cannot be merged~\cite{adsenseAccess}. 
An AdSense account cannot be transferred to another individual~\cite{adsenseFaq}, but two or more AdSense accounts with different \publisherIDs can co-exist on the same website~\cite{adsenseShare}. 
These other \publisherIDs can belong to a business partner, contributing authors, or even third-parties. 

\subsection{Google Tag Manager (GTM)}
\label{sec:googleTagManager}

\emph{Google Tag Manager} (GTM) is a service for web administrators to manage code snippets (called \emph{Tags} - provided by third-parties to integrate their respective service \eg/ analytics, marketing, support) in their website. 
GTM provides an interface for publishers to handle such code snippets, which uses an abstraction called container that needs to be installed in the website by inserting its own snippet~\cite{farney2016google}. 
A container is uniquely identified by a \containerID, formatted as \texttt{GTM-XXXXXX}.
One GTM account can create and manage more than one containers. 
Usually, a GTM account represents the topmost level of organization and, typically, an organization uses a single GTM account~\cite{gtm}. 
Thus, containers are not bound to a domain or a website, and with the appropriate configuration, the same container can be used in multiple websites~\cite{containers}. 

\subsection{Google Analytics}
\label{sec:googleAnalytics}

\emph{Google Analytics} is a service to track and report website traffic. 
The service revolves around Properties, which contain the reports and traffic data for one or more websites or applications. 
There are two types: (i) \emph{Google Analytics 4} Properties that are identified by a \measurementID, which follows the format \texttt{G-XXXXXXX}, and (ii) \emph{Universal Analytics} (the older version of properties~\cite{universalAnalytics}) that are uniquely identified by a \trackingID, formatted as \texttt{UA-000000-1}. 
When a user creates a Google Analytics account, a unique identifier is created that acts as a prefix of \trackingIDs (\ie/ the first set of numbers).
Consequently, the \trackingID which is included in the code snippet indicates which account data is sent to~\cite{trackingID}. 
The suffix of a \trackingID represents the property that data is sent to. 
A website publisher that owns more than one website is able to associate a single property with all of these websites.
\section{Methodology}
\label{sec:methodology}

\begin{table}[t] 
\footnotesize
\caption{Detected \identifiers and their origin.}
\vspace{-0.1cm}
    \centering
    \begin{tabular}{lrrr||rrr}
    \toprule
        & \textbf{Unique}   & \textbf{Unique} & \textbf{\% of} & & & \\
        \textbf{Description}  & \textbf{IDs} & \textbf{URLs} & \textbf{sites} & \textbf{HTML} & \textbf{Reqs} & \textbf{Cookies} \\
    \midrule
        \publisherIDs   & 71,745  & 87.273  & 10.05\% & 99.4\% & 77.6\%   & 0.00\% \\
        \trackingIDs    & 485,405 & 451.498 & 52.02\% & 76.9\% & 94.3\%  & 22.5\% \\
        \measurementIDs & 47,087  & 47,606  & 5.48\% & 96.0\% & 91.2\%  & 0.36\% \\
        \containerIDs   & 193,693 & 179,114 & 20.64\% & 99.7\% & 93.0\%   & 0.01\% \\
    \bottomrule
    \end{tabular}
    \label{tab:identifiers}
\end{table}

\subsection{Crawling Methods}
\label{sec:crawling}

To detect websites operated by the same entity, we search for the identifiers of the respective services described in Section~\ref{sec:background}.
Specifically, we develop a Puppeteer-based crawler that instruments instances of the Chromium Browser.
Using these instances, we crawl with clean state the landing page of the top 962K websites of the Tranco list, which aggregates the ranks from the lists provided by Alexa, Umbrella, and Majestic, from 16/3/2021 to 14/4/2021\cite{trancoList}.
This list is formed based on techniques which enable list stability, facilitate reproducibility, and protect against adversarial manipulation.
The implementation of our crawler is publicly available~\cite{openSourceCode}.
When our crawler visits a website, it waits until the page has completely loaded and for an additional 5-sec period to ensure that all programmatically purchased ads (via Real Time Bidding (RTB)) have been rendered.
Then, it stores the HTML of the page, a cookie-jar and the HTTP(S) requests performed during the website visit.
We capture all requests passively in a read-only fashion without mutating or intercepting them.
This ensures that the behavior of the website is not affected by our crawler.
To collect the HTML of the website, we utilize the Chrome DevTools Protocol~\cite{chromeDevTools}.
This way, we ensure that we capture not only the actual HTML code but also the documents, styles or code fetched by iFrames or code snippets.
Our crawler visits 962K websites of the Tranco list (\textbf{\trancoCrawl}) from 15-27/4/2021 and collects 415GB of data.
In 93,817 websites (9.75\%) the crawling process failed due to timeouts or site inaccessibility.
Overall, we detect 525,493 websites, with at least one of the identifiers discussed in Section~\ref{sec:background}.
Ethical concerns regarding the crawling process and collected data are addressed in Appendix~\ref{sec:ethics}.

\subsection{Detecting Identifiers}
\label{sec:detecting-ids}

We detect the Google Identifiers described in Section~\ref{sec:background} by performing an offline analysis on the collected data.
Specifically, using regular expressions\footnote{pub-[0-9]\{9,\}, UA-[0-9]\{4,\}-[0-9]+, G-[A-Z0-9]\{7,\} and GTM-[A-Z0-9]\{6,\}}, we search for these identifiers inside the page content, HTTP(S) requests and stored cookies.
Then, we remove false positives using a combination of data-filtering techniques.

First, using the dictionary of GNU Aspell~\cite{aspell}, an open-source spell-checking tool, we remove values which are words of the English dictionary, but match the suffix of regular expressions (\eg/ \texttt{G-BACKPACK}).
Using this technique, we were able to remove $\sim$1,500 distinct false positive values, which were found in $\sim$5,000 unique websites.
Then, we remove false positives using a list of common keywords.
This list was generated by manually inspecting over 10,000 values that satisfy the regular expressions, and investigating whether they are actually used as identifiers.
Our keyword list contains over 1,250 values which were filtered out (\eg/ \texttt{G-APRIL2020}).

As in Table~\ref{tab:identifiers}, we find that $\sim$10\% of the most popular websites monetize their content through an AdSense account (\ie/ \publisherID), $\sim$52\% use Google Analytics to track their traffic, and $\sim$20\% use the Google Tag Manager for easier management of code snippets.
Moreover, for some services, we observe that there are more domains than \identifiers.
This suggests that some identifiers are being re-used in more than one website.
Additionally, we examine the source of information for each type of identifier.
Specifically, we investigate whether the identifiers can be found in the HTML code of a website, its outgoing network traffic, or in the cookies set by either the first-party or various third-parties.
As shown in Table~\ref{tab:identifiers}, regardless of the type of identifier, the majority of them can be found in both the HTML code of the website and the HTTP(S) requests.
This result is inline with the official guidelines for using Tags~\cite{tags}.
This indicates that the detected identifiers are not only valid, but since they are sent to the respective Google services, they are in use.
Finally, we find that only \trackingIDs are commonly found in cookies.

\begin{figure}[t]
    \centering
    \includegraphics[width=0.5\columnwidth]{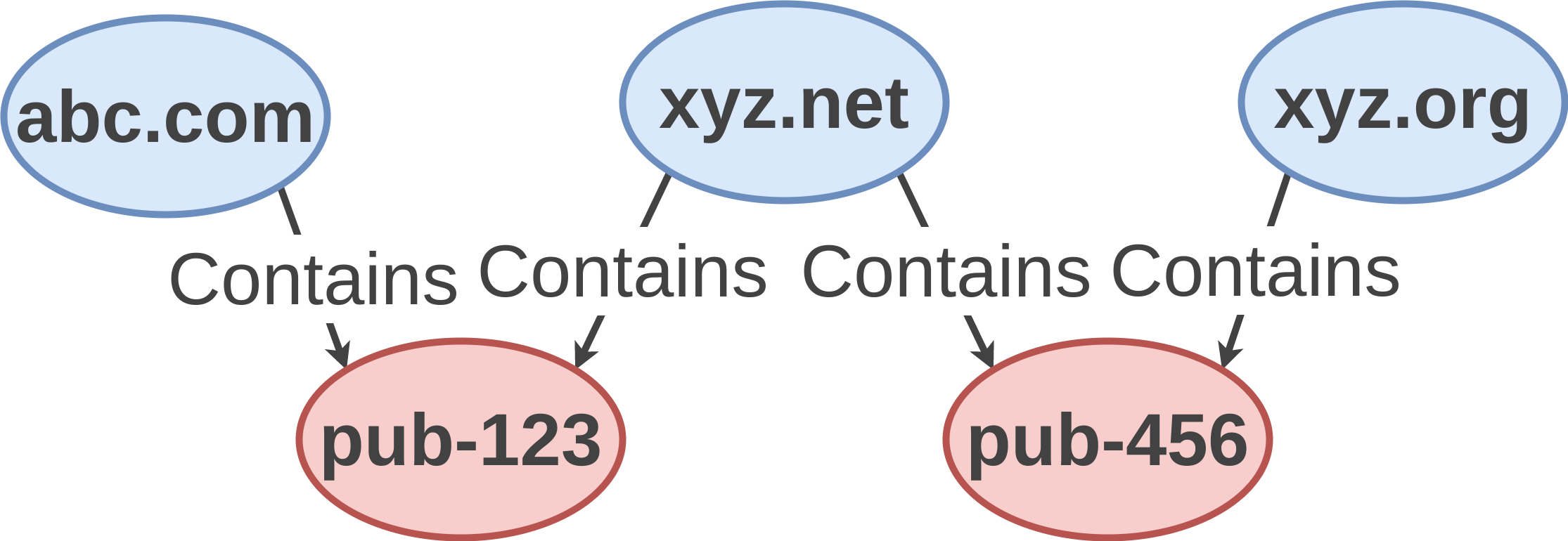}
    \vspace{-0.1cm}
    \caption{Example of a \publisherID bipartite graph. Blue nodes represent websites and red nodes represent \publisherIDs. A directed edge in these bipartite graphs indicates that the website contains the respective identifier.}
    \label{fig:bipartitegraph}
\end{figure}
\begin{figure}[t]
    \centering
    \includegraphics[width=0.78\columnwidth]{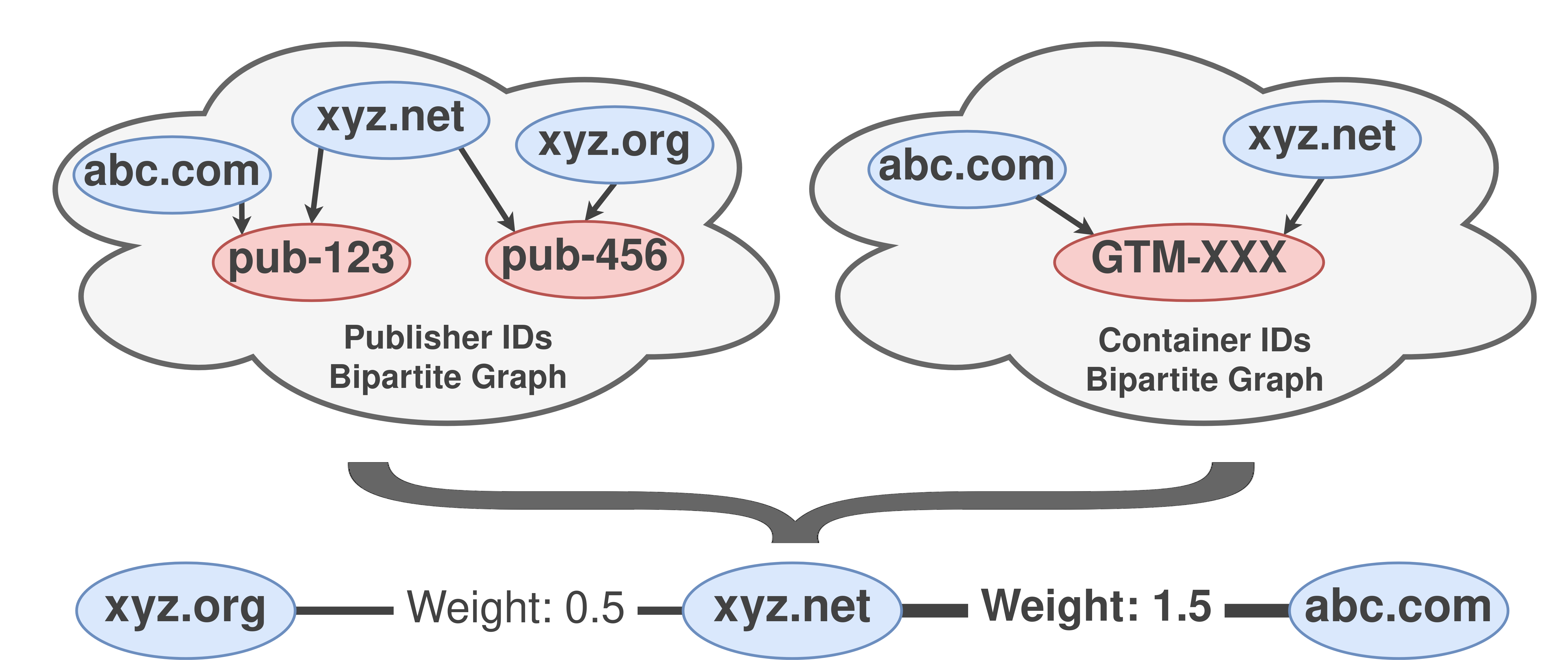}
    \vspace{-0.1cm}
    \caption{Example of \metagraph construction. Websites that share an identifier, share an undirected edge in the resulting graph. The weight of the edge rises proportionally with the number of common \identifiers.}
    \label{fig:hypergraph}
    \vspace{-0.2cm}
\end{figure}

\subsection{Bipartite graphs}
\label{sec:bipartiteGraphs}

Using the detected \identifiers, we construct a bipartite graph for each of the respective types of identifiers.
In these graphs, the nodes are either websites or identifiers.
Whenever a website contains an identifier, we introduce a directed edge from the respective website node to the respective identifier node.
\trackingIDs and \measurementIDs are placed into the same graph since they represent the same service and have similar functionality.
Thus, we create three bipartite graphs.
Moreover, for \trackingIDs, we focus only on the prefix, which refers to the account number, as discussed in Section~\ref{sec:googleAnalytics}.
Figure~\ref{fig:bipartitegraph} illustrates an example of a small \publisherID bipartite graph.

\subsection{\Metagraph}
\label{sec:metagraph}

We also form a \metagraph based on the three bipartite graphs of different \identifiers.
The \metagraph contains nodes only for websites and represents the relationships between websites.
Whenever two websites share an identifier, we introduce an undirected meta-edge between the two respective website nodes.
The more identifiers two websites share, the greater the weight of the connecting edge.
Each shared identifier increases the weight of the meta-edge by $\frac{1}{n}$, where $n$ is the total number of distinct identifiers of this type found in more than one website.
A larger edge weight between two websites implies greater confidence that they are indeed operated and monetized by the same entity.
Figure~\ref{fig:hypergraph} illustrates an example of how the \metagraph is constructed.
The code to construct both the bipartite graphs and the \metagraph is publicly available~\cite{openSourceData}.

\begin{figure}[t]
    \centering
    \includegraphics[width=0.7\columnwidth]{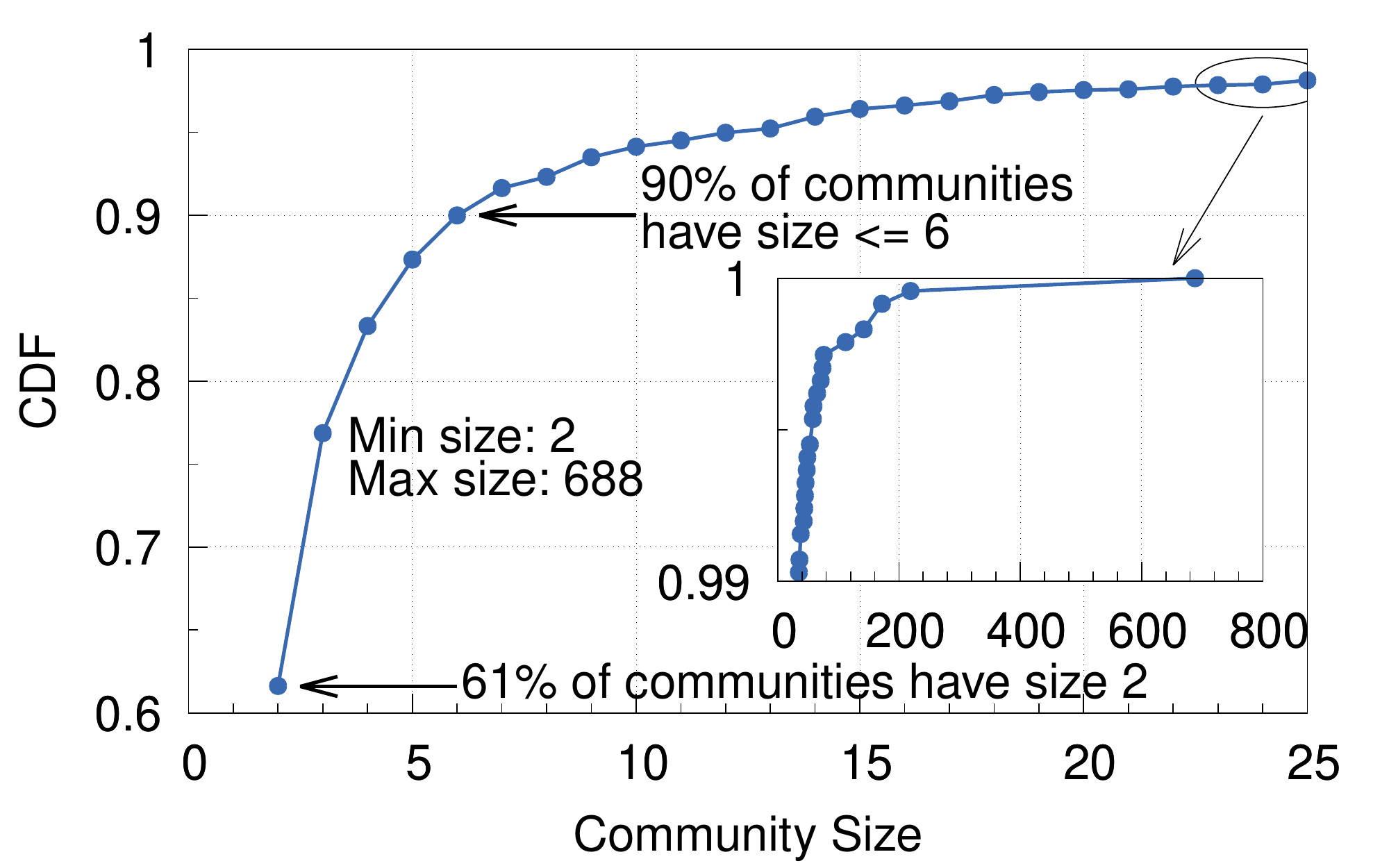}
    \vspace{-.1cm}
    \caption{Distribution of Girvan-Newman community sizes in the \metagraph. Over 60\% of the communities are made of just two sites.}
    \vspace{-.3cm}
    \label{fig:hypergraphCommunitySizes}
\end{figure}

\begin{figure*}[t]
    \begin{minipage}[t]{0.323\textwidth}
        \includegraphics[width=0.9\columnwidth]{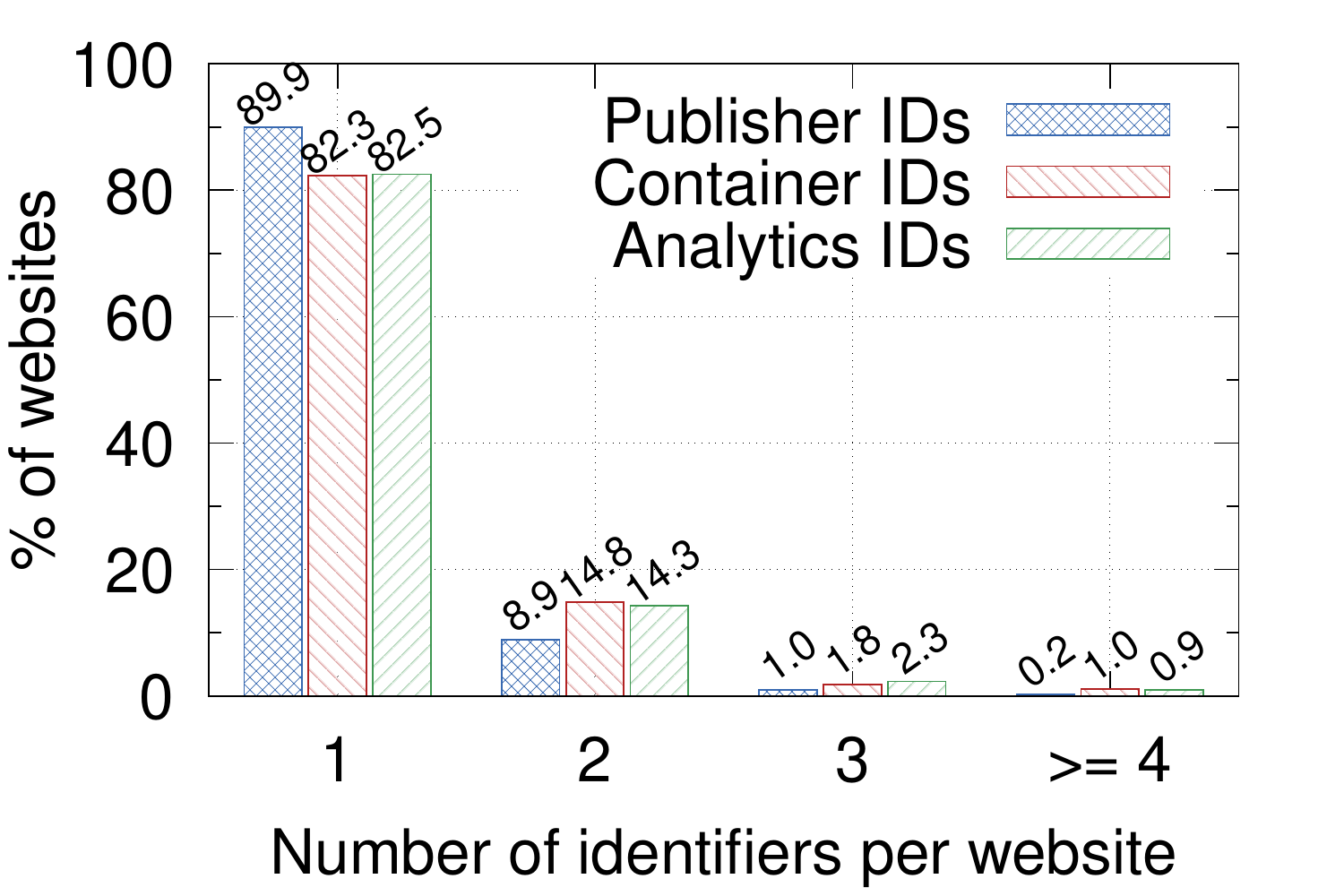}
        \caption{Distribution of \identifiers per website. Most websites contain only one identifier.}
        \label{fig:containedIDs}
    \end{minipage}
    \hfill
    \centering
    \begin{minipage}[t]{0.323\textwidth}
        \centering
        \includegraphics[width=0.9\columnwidth]{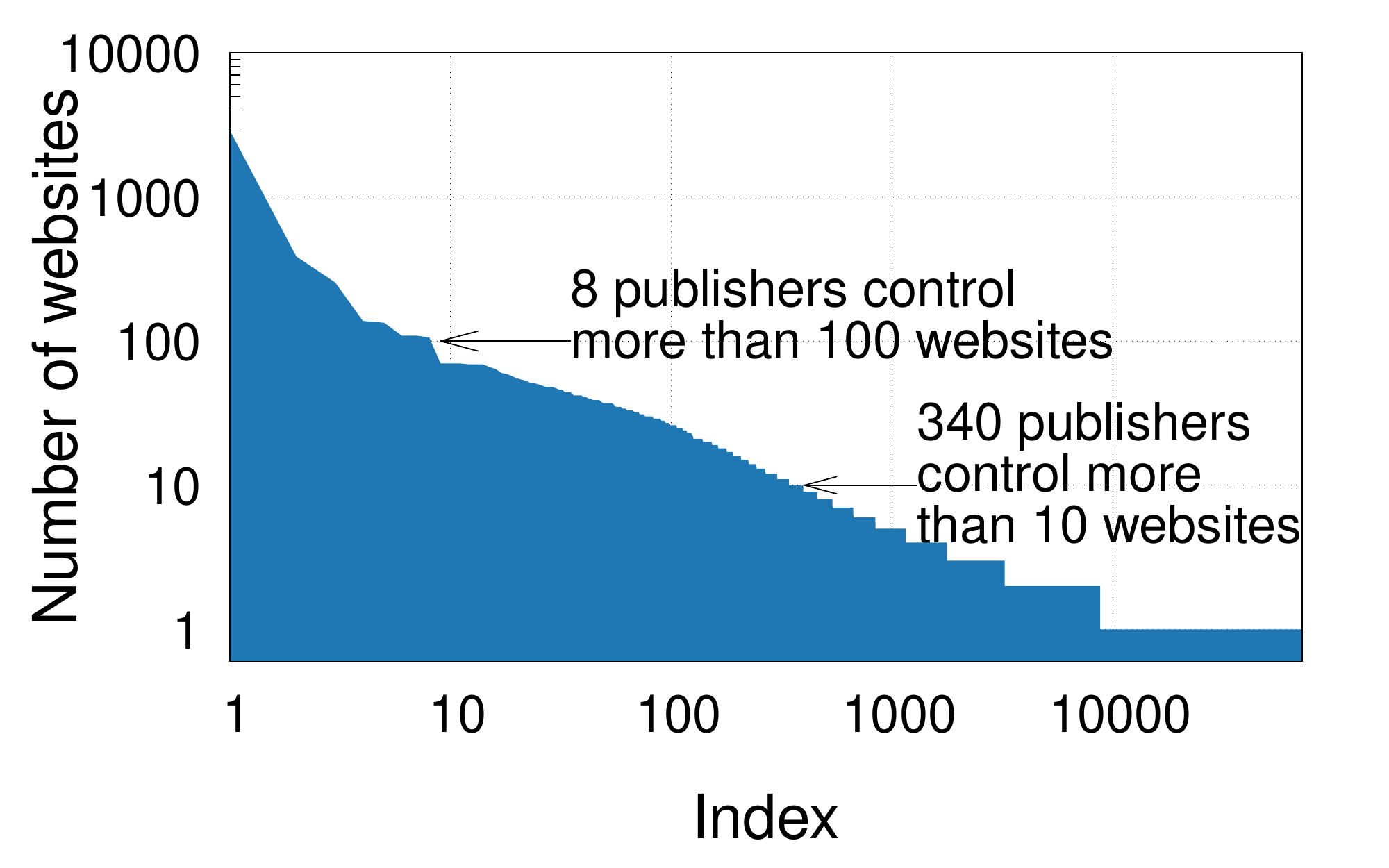}\
        \caption{Number of websites monetized by each publisher. Most publishers ($\sim$63,000) appear in only one site.}
        \label{fig:publisherSize}
    \end{minipage}
    \hfill
    \begin{minipage}[t]{0.323\textwidth}
        \centering
        \includegraphics[width=0.9\columnwidth]{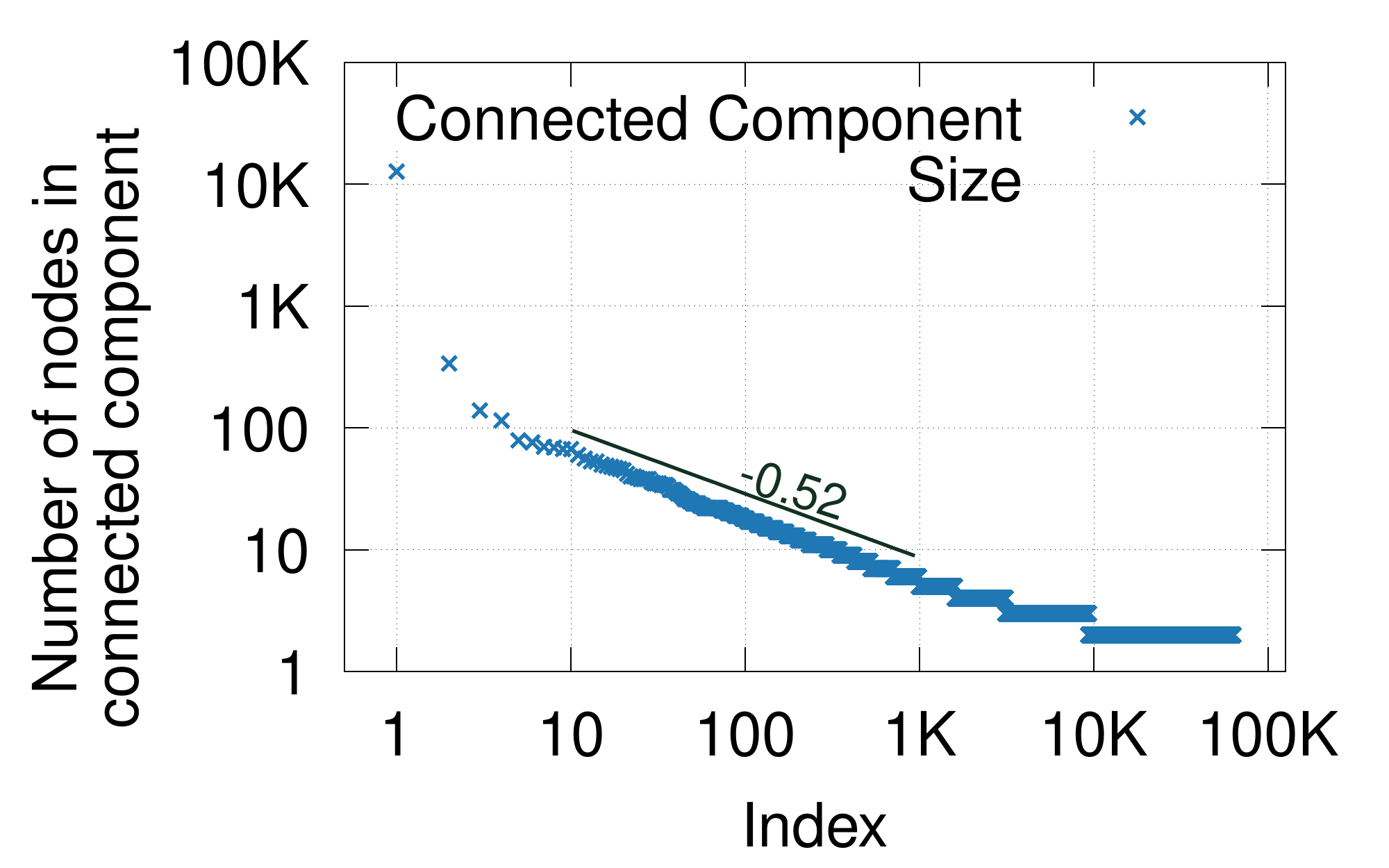}
        \caption{Connected components sorted by size (number of nodes) for the \publisherID bipartite graph.}
        \vspace{-.4cm}
        \label{fig:connectedComponents}
    \end{minipage}
\end{figure*}

\subsection{\Metagraph Validation}
\label{sec:metagraph-validation}

We hypothesize that the \metagraph constructed by following the steps above can lead to clusters of websites operated by the same entity. 
Meta-edges reflect the actual relationships between websites, thus the ones operated by the same entity should form a strongly connected community.
Since the \metagraph combines information from multiple service (\ie/ bipartite graphs), it provides us with greater confidence about the actual relationships of the websites.

We find that there are some outlying cases where communities consist of thousands of websites. 
After manual investigation, we conclude that these communities are formed due to \emph{intermediary publishing partners}. 
These publishers provide services for content creators to monetize their content or improve their website traffic and require that websites integrate the partner's identifiers. 
To focus our analysis on an enhanced level of granularity, we ignore such publishers for the time being.
This allows us to study more detailed cases of website administrators and results in a \metagraph that contains $\sim$127,000 nodes and $\sim$2,885,000 edges. 
We discuss intermediate publishing partners in later sections. 

To find the websites operated and monetized by the same entity with high confidence, we perform edge pruning, thus removing noise. 
Specifically, we remove edges that do not belong to the top 5\%, when ranked by weight.
We choose this threshold based on empirical analysis. 
This way, we ensure that there are limited false positives, \ie/ websites that are wrongfully added to a community because of a typographical error in their source code, or older identifiers. 
After the edge pruning, dangling nodes are also removed from the graph as they do not provide any additional information. 
To further explore this graph, we execute the Girvan–Newman community detection algorithm~\cite{girvan2002community}. 
In fact, we compare our methodology with the one described in~\cite{cangialosi2016measurement}, where the authors apply the Louvain method~\cite{blondel2008fast} to the connected components extracted from their bipartite graph. 
Specifically, we manually examine and evaluate 40 distinct communities, which can be found through both methodologies, consisting of 215 unique websites. 
We find that applying the Girvan-Newman algorithm, after edge-pruning, results in better communities in 42.5\% of the cases, in exactly the same communities 37.5\% of the cases, and in worse communities in 20\% of the cases. 
Consequently, the Girvan–Newman algorithm results in higher quality communities at the expense of a higher computational cost.

Finally, we compare the communities our methodology detects with the (publicly available) communities detected in~\cite{cangialosi2016measurement}. 
We acknowledge that the comparison is difficult to achieve because the two studies focus on different websites, in a different time period (\ie/ 5-year ago: a big portion of the websites are no longer active and cannot be evaluated) and in a very dynamic environment like the Web. 
We compare only communities that contain websites, all of which have been crawled in our \trancoCrawl. 
In total, we manually evaluated and found 12 communities with results similar to ours and 15 cases where the methodology of~\cite{cangialosi2016measurement} fails and places websites which are operated by the same legal entity into different communities. 
Using our methodology we detect 2,369 communities, formed by $\sim$11,000 distinct websites. 
The distribution of community sizes (Figure~\ref{fig:hypergraphCommunitySizes}) shows that the majority of them are small (\ie/ <6 websites). 
Indeed, 61\% of the communities are pairs of websites, indicating that the median publisher operates just 2 websites.

\begin{figure*}[t]
    \centering
    \begin{minipage}[t]{0.235\textwidth}
    \centering
    \includegraphics[width=1.12\columnwidth]{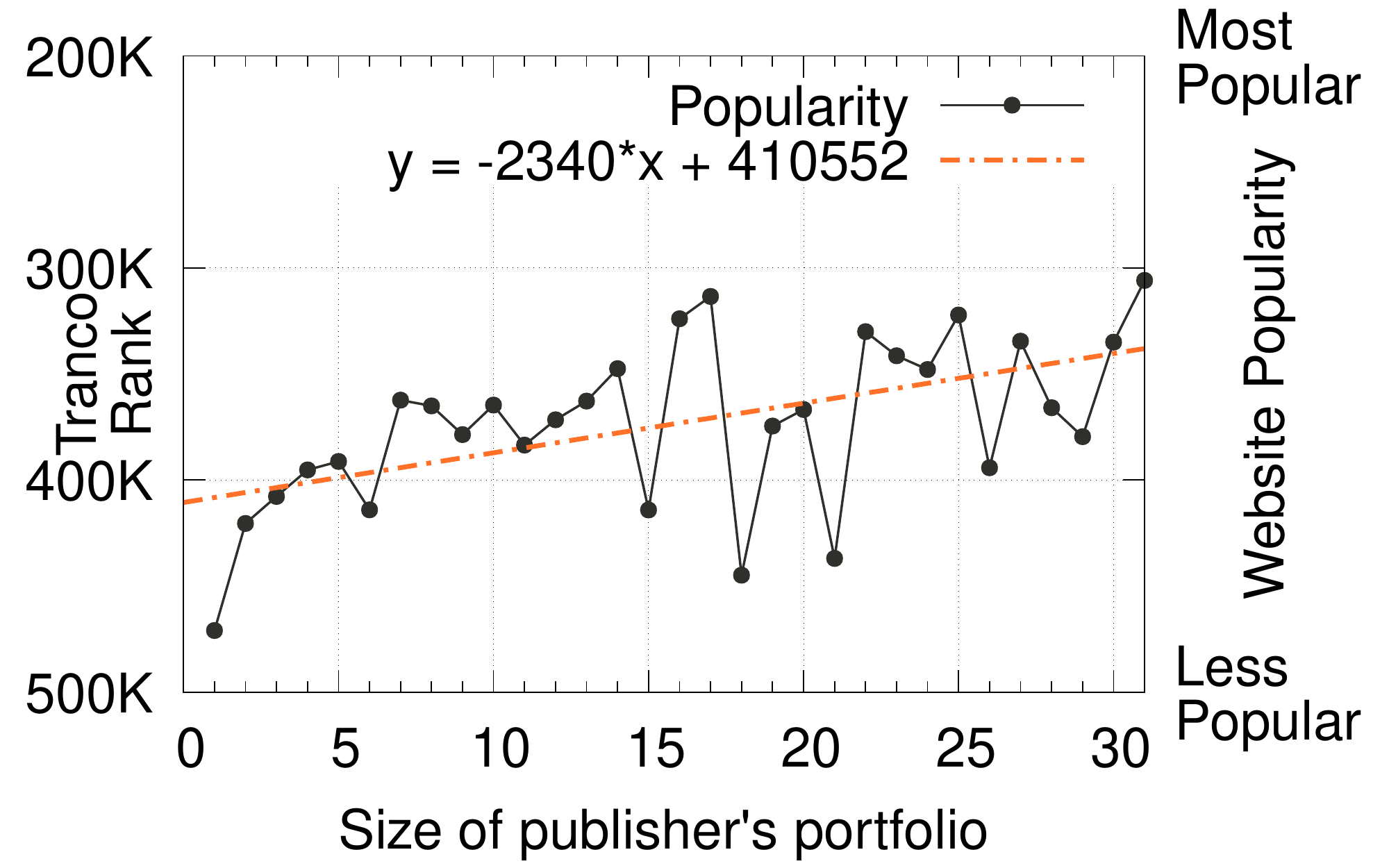}
    \caption{Average popularity of publishers (based on ranking of sites) vs. publisher size.}
    \label{fig:ranksOfPublishers}
    \end{minipage}
    \hfill
    \begin{minipage}[t]{0.235\textwidth}
        \centering
        \includegraphics[width=1.12\textwidth]{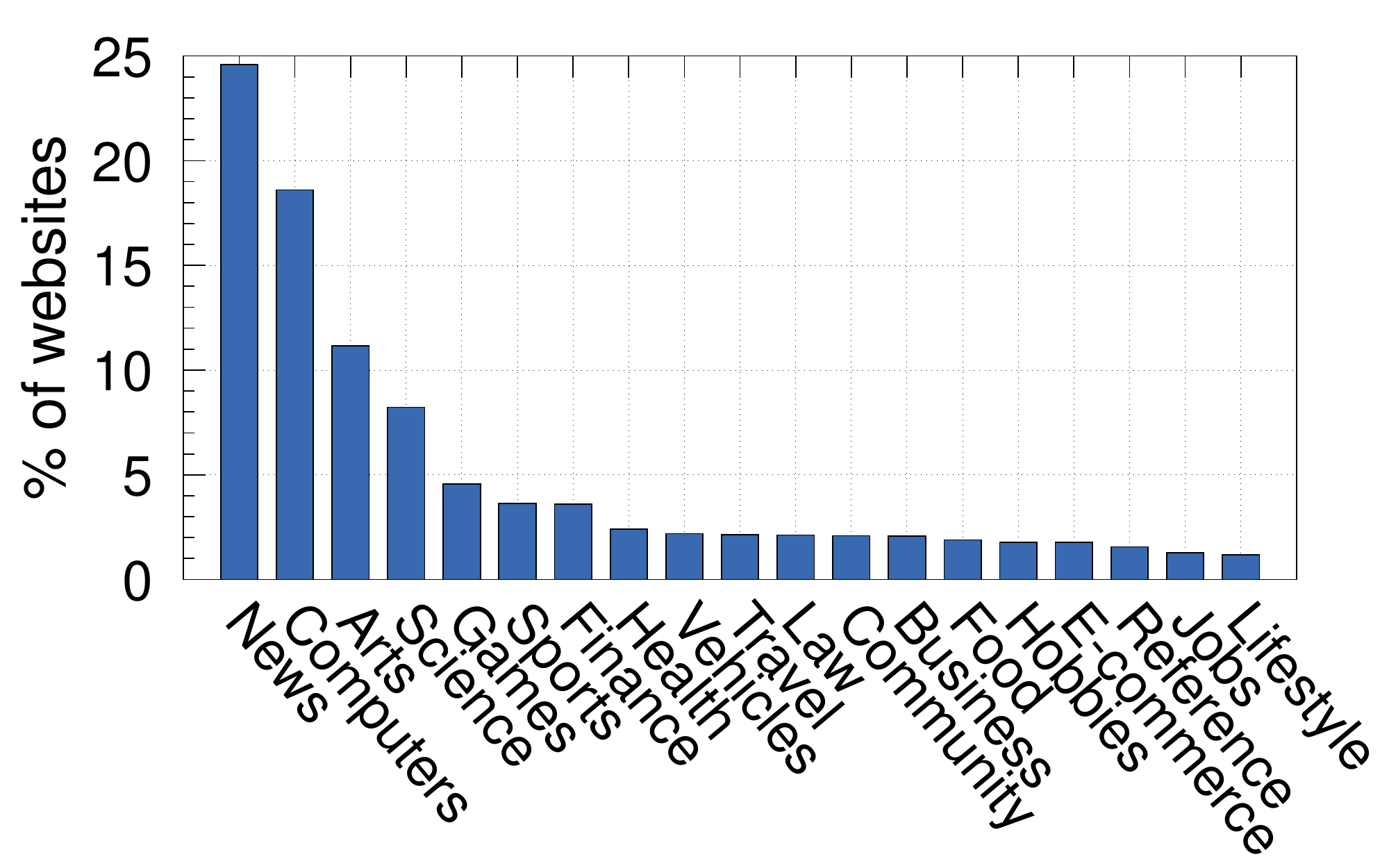}
        \caption{Distribution of categories for websites with \publisherID.}
        \label{fig:categoriesDistribution}
    \end{minipage}
    \hfill
    \begin{minipage}[t]{0.235\textwidth}
        \includegraphics[width=1.15\textwidth]{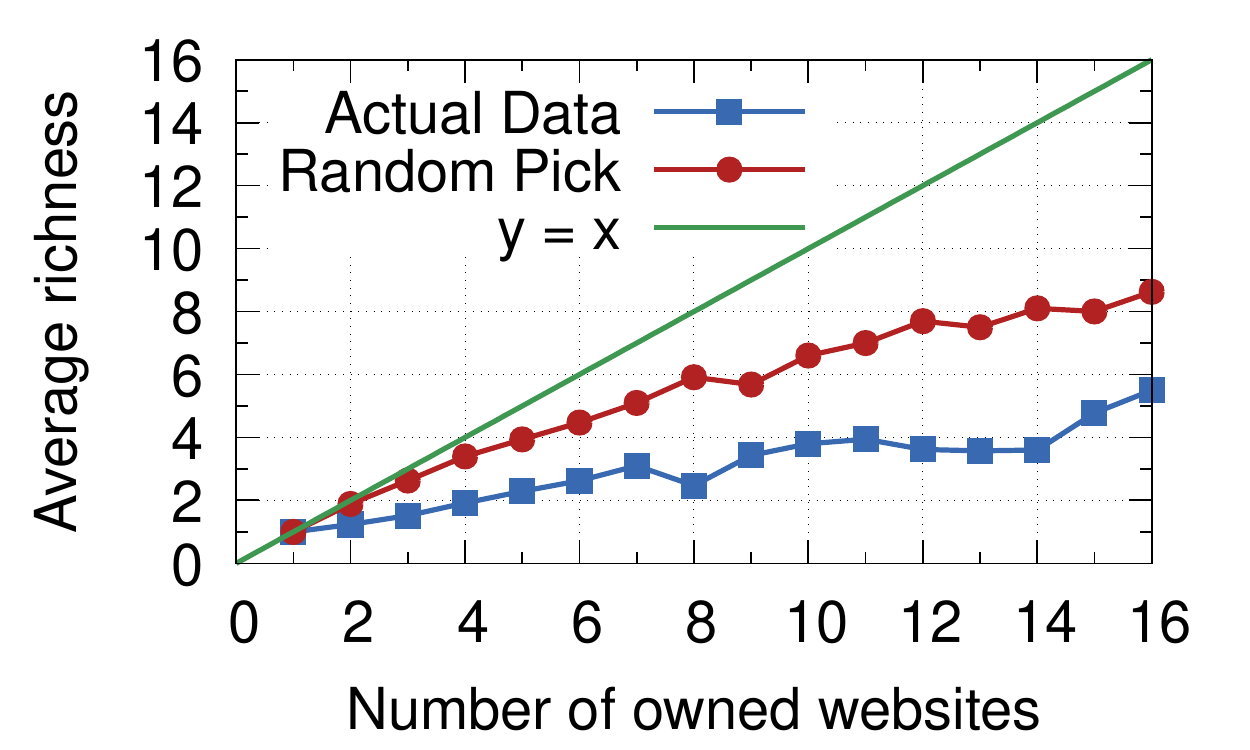}
        \caption{Poisson sampling experiment for site categories}
        \label{fig:randomChoice}
    \end{minipage}
    \hfill
    \begin{minipage}[t]{0.235\textwidth}
    \includegraphics[width=1.14\textwidth]{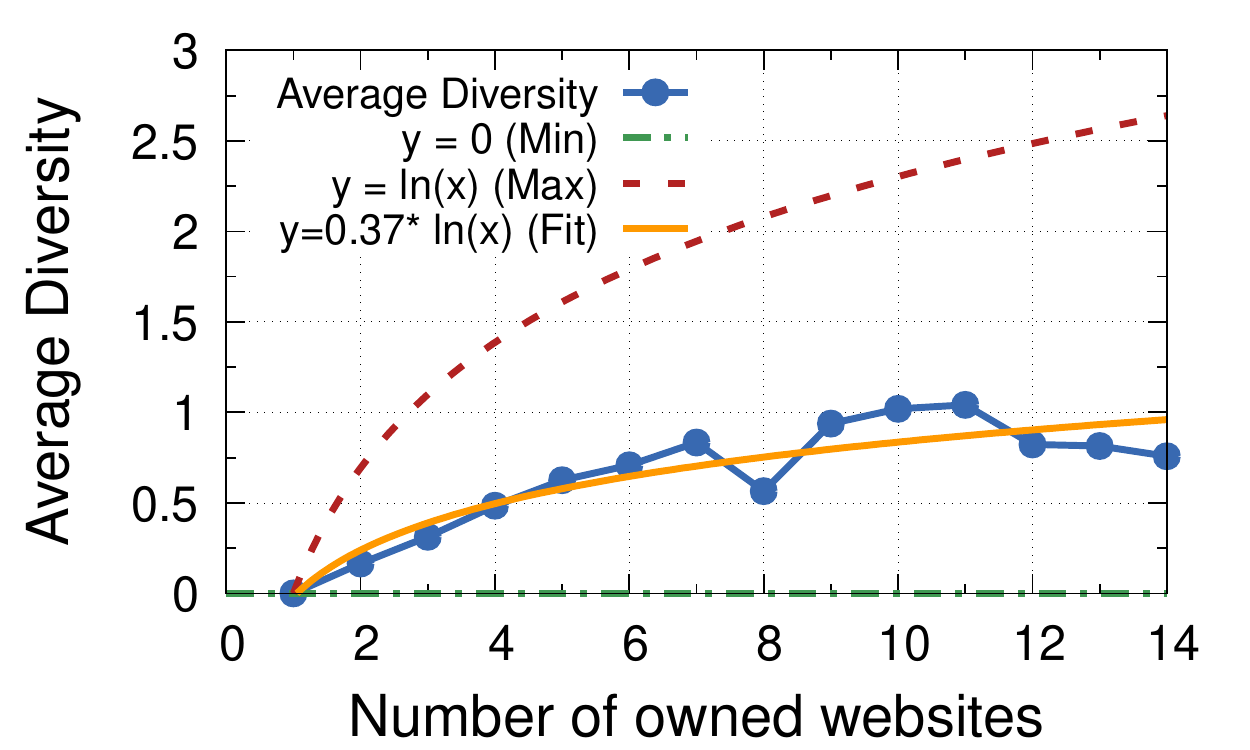}
    \caption{Shannon's Diversity Index for website categories in detected communities.}
    \label{fig:shannonIndex}
    \end{minipage}
\end{figure*}
\section{Analysis of Bipartite Graphs}
\label{sec:analysis}

\subsection{Contained \identifiers}
\label{sec:ids-in-pages}

First, we study how organizations or publishers use identifiers necessary for Google services.
For this, we measure the number of unique identifiers found in each website of our dataset.
In Figure~\ref{fig:containedIDs}, we plot the portion of websites in our data that contain a certain number of identifiers.
Around 82-83\% of the websites contain only one \containerID or \analyticsID, and about 90\% of websites have only one \publisherID.
This indicates that the majority of organizations prefer to use the simplest and most straightforward configuration  of services in their websites, where they use a single identifier to achieve their goal, be it monetization or traffic measuring.
Most importantly, in the case of \publisherIDs, it indicates that the majority of websites have a single contributing author and that revenue is not shared.
This is contrasting the small portion of websites (less than 3.2\% in all cases), with 3 or more identifiers, which indicate multiple collaborating authors, with their own \publisherIDs, contributing to a website.
Surprisingly, we see a small number of websites with an extremely large number of identifiers.
For instance, we find that \emph{prykoly.ru} contains 94 \trackingIDs, while \emph{www.pps.net}, a website for public schools in Portland, contains 88 IDs hard-coded in the JavaScript code and the correct identifier is selected based on the page that is visited.
To further investigate this an abnormal behavior, we lookup these websites in the \emph{VirusTotal}\cite{virustotal} and \emph{Sucuri}\cite{sucuri} security services for malicious content.
Sucuri reports\cite{sucuriReport} that prykoly.ru contains a known JavaScript malware associated with a back-link purchase service, called Sape.
For pps.net, Virustotal reports\cite{virustotalReport} that there are 4 detected files that communicate with this domain.
In total, we find 67 distinct websites with over 40 \identifiers of any type.
This preliminary analysis suggests that cases with numerous \identifiers in a website might imply abnormal or even malicious behavior.
This observation, though interesting, is considered as out of scope for this work and left for future research.

\subsection{Publisher Size}
\label{sec:publisher-size}

Next, we explore the amount of websites that publishers manage and monetize.
For each publisher, we measure the number of websites in which they place their \identifiers.
Analysis from now on, is performed on distinct domains of landing pages, and not distinct domains in the Tranco list.
Specifically, if two different domains in the Tranco list redirect to the same domain, we measure this website only once towards the size of the respective publisher.
In Figure~\ref{fig:publisherSize}, we plot, in descending order, the number of websites monetized by each unique \publisherID in our data.
We show that the great majority of publishers (up to 87.8\%), monetize traffic from a single site.
On the other hand, we find 340 publishers monetizing traffic from more than 10 websites each.

We observe some ``mega-publishers'' that can be found in hundreds or even thousands of websites.
Indeed, the top 10 publishers in our data can be found in a total of more than 4,200 websites.
We observe similar behavior in \containerIDs and \analyticsID, where we find that the top 10 identifiers can be found in a total of 4,245 and 6,795 websites, respectively.
To verify this finding, we explore the connectivity of the three bipartite graphs (described in Section~\ref{sec:bipartiteGraphs}) and generate a list of connected components in each graph.
In Figure~\ref{fig:connectedComponents} we plot, in decreasing order using a log-log scale, the number of nodes in each connected component of the \publisherID bipartite graph.
We see that the distribution of connected component sizes can fit a power-law with a cutoff, an anomaly due to the intermediary publishers mentioned earlier.
By applying appropriate statistical tests~\cite{clauset2009power}, we find that the distribution is indeed heavy-tailed with power law being a better fit than the exponential distribution (loglikelihood ratio test with $p = 1.6e ^{-8}$).
We find similar results for the \containerIDs and \analyticsIDs bipartite graphs but exclude them for brevity.
This verifies our finding that there are only a few publishers monetizing traffic from a very large number of websites, while the majority of publishers operate one website.

We attribute this behavior to the existence of \emph{intermediary publishing partners}~\cite{googlePartners}.
These are third-party services which provide services for content creators to readily deliver their content, effortlessly monetize it, optimize their revenue and deliver better experiences to users.
Publishers are required to integrate the service's identifiers in their websites so that the service can monitor traffic and user behavior, and deliver ads.
By examining requests towards third-party services, we successfully identified multiple such ``mega-publishers'', including \textit{Ezoic}, \textit{optAd360}, \textit{Blogger} and \textit{ProjectAgora}.
Specifically, \textit{Blogger} provides publishers with AdSense gadgets, which can be used to display ads in a blog, without taking any percentage of earnings~\cite{bloggerAds}.
At the moment of writing, the \emph{PublicWWW} service~\cite{publicWWW} reports that Blogger's \publisherID can be found in more than 364,000 websites.

\subsection{Monetizer Popularity}
\label{sec:website-popularity}

Next, we explore if there is an association between the popularity of website and the size of the publishers.
First, we group together websites that share the same \publisherID, meaning that there is a single account responsible for their monetization.
For each such account (\ie/ publisher), we compute its popularity as the average rank of the websites it operates, based on the Tranco list.
In Figure~\ref{fig:ranksOfPublishers}, we plot this average popularity of publishers, for a given size of publishers.
We show that the average website popularity (y-axis) increases (\ie/ Tranco rank decreases), as the number of monetized websites increases (x-axis).
The average popularity subsections have also been fitted with a straight line (the negative slope is from the reversed y-axis to indicate increased popularity) indicating a clear trend with a $R^2=0.28$.
We observe similar behavior when plotting the median popularity, indicating that there is no skewness in the distribution.

As a result, independent publishers who generate revenue from a single website, tend to monetize less popular websites.
On the other hand, publishers that manage multiple websites, usually manage the most popular ones.
Consequently, in a classic case of \emph{rich getting richer}, big publishers who operate dozens of websites, not only claim a bigger share of the market and generate a bigger revenue, but they are also able to improve their reputation and attract more visitors.
The increased popularity of some websites can also be credited to the intermediary publishers mentioned earlier.

\begin{figure*}[t]
    \centering
    \begin{minipage}[t]{0.323\textwidth}
        \centering
        \includegraphics[width=1.02\textwidth]{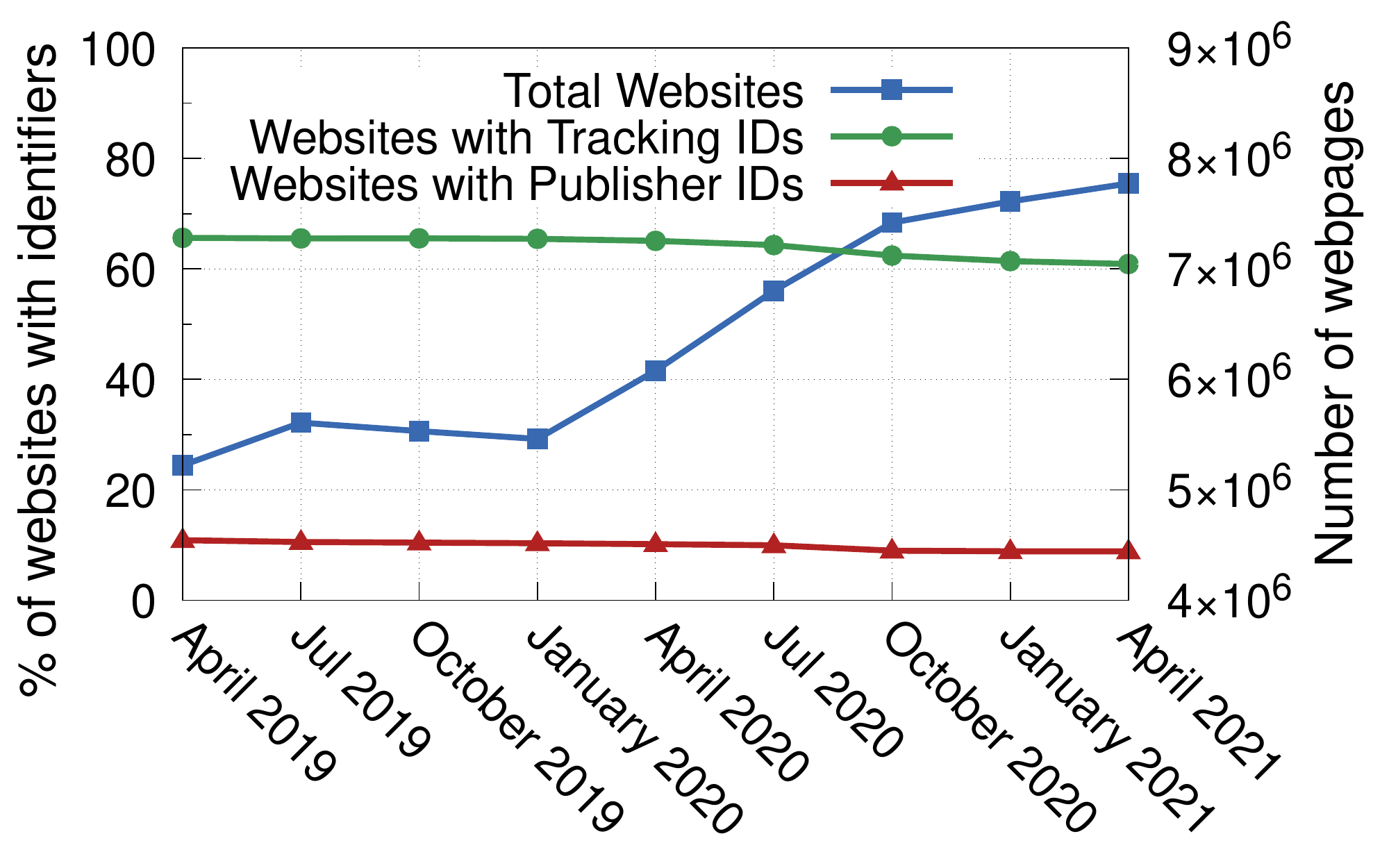}
        \caption{Total websites crawled and portion of websites that contain \publisherID and \trackingIDs.}
        \label{fig:httpArchiveDataset}
    \end{minipage}
    \hfill
    \begin{minipage}[t]{0.323\textwidth}
        \centering
        \includegraphics[width=1.02\textwidth]{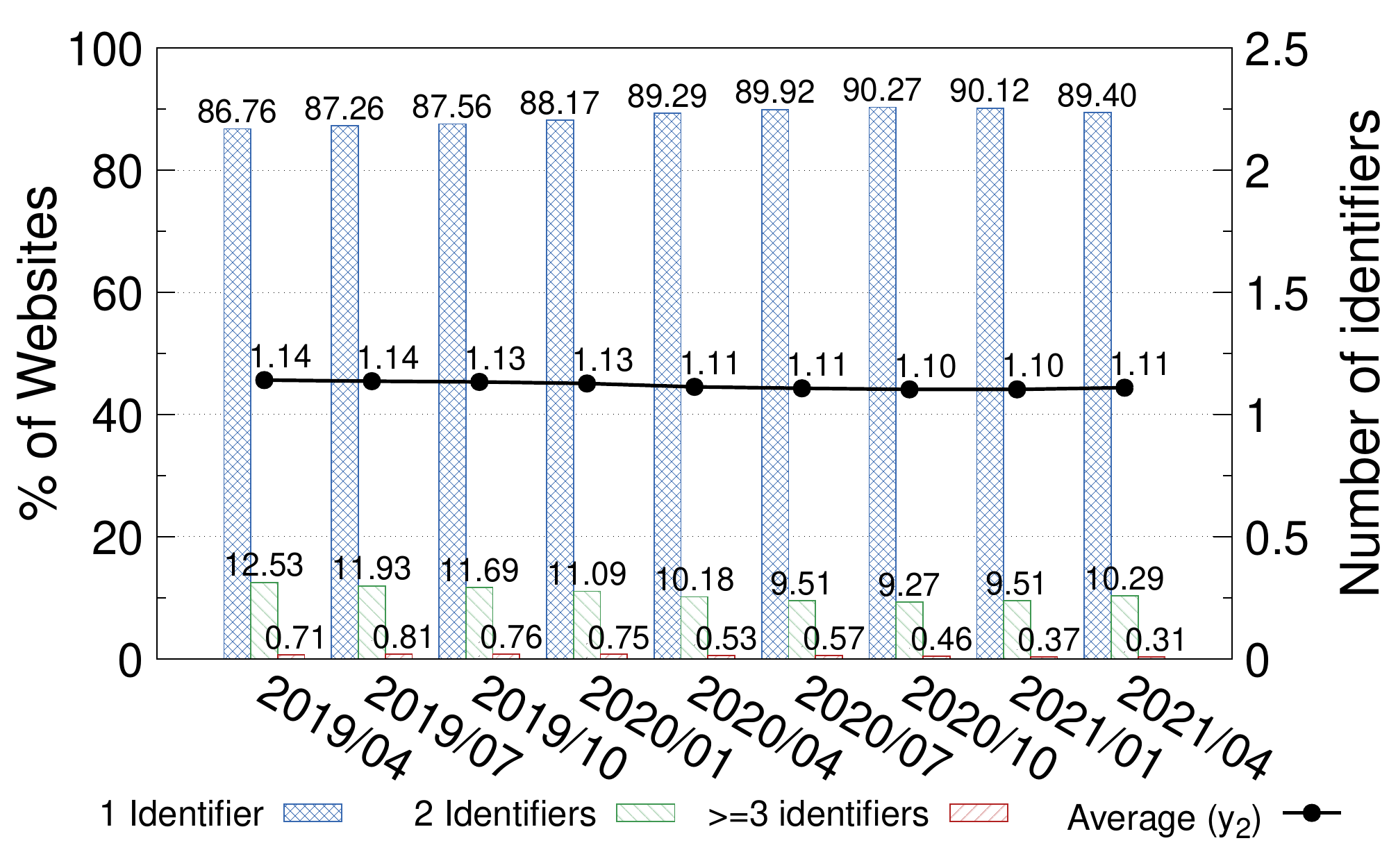}
        \caption{Distribution of \publisherIDs per website through time.
        Almost 90\% of websites in all snapshots contain only one \publisherID.}
        \label{fig:historyContainedIDs}
    \end{minipage}
    \hfill
    \begin{minipage}[t]{0.323\textwidth}
        \centering
        \includegraphics[width=1.02\textwidth]{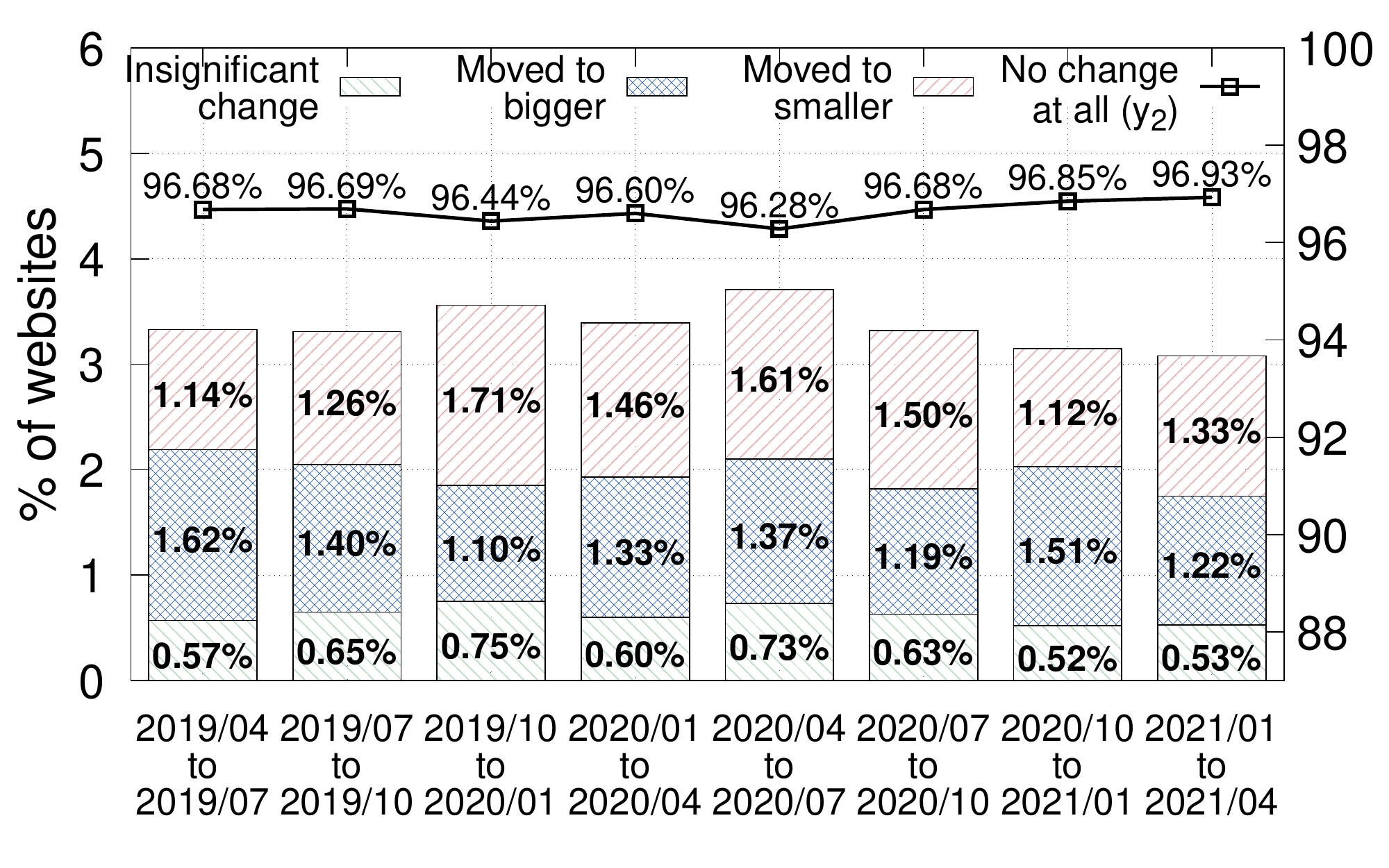}
        \caption{Transition of websites between publishers through time.}
        \label{fig:historyMovement}
    \end{minipage}
    \vspace{-0.4cm}
\end{figure*}

\subsection{Website Categories}
\label{sec:publisherid-categories}

Manual inspection of communities (Section~\ref{sec:metagraph-validation}) revealed that most operators tend to manage websites with similar content.
To investigate this further, we retrieve (if available) from \emph{SimilarWeb}~\cite{similarWeb} the category of each website with a \publisherID in our dataset.
Figure~\ref{fig:categoriesDistribution} illustrates the distribution of the categories we retrieved from over 23,000 websites with a \publisherID.
Websites with no category information are excluded from the analysis.
We see a preference towards ``News and Media'' websites (24.5\%), followed by websites related to ``Computers, Electronics and Technology'' (18.6\%), ``Arts'' (11\%) and ``Science'' (8\%).

Next, we investigate if there is a preference in the types of websites a publisher monetizes, or if their portfolio is usually random.
As a first step, we perform a Poisson Sampling experiment to construct a scenario where publishers monetize websites based on their categories' measured popularity in the data.
For this sampling, we perform the following steps.
For a given size of publisher (\ie/ number of websites they operate and monetize), we randomly select websites from our data.
For example, if the size of a publisher is 10, we randomly select 10 websites from the 23K websites with a category.
However, this selection is biased, based on the prior probability of a website appearing due to its category.
Thus, for example, a ``News and Media'' website has almost 1/4 chance to be selected in any publisher.
We perform this process for all publishers.
Then, for each publisher, we compute the number of unique website categories they have in their control (\ie/ \emph{richness}).

Figure~\ref{fig:randomChoice} plots the richness distribution of the 
observed (or ``actual'') data and the Poisson-Sampling experiment data.
The $y=x$ line represents the case of a uniform distribution, where all of a publisher's websites come from different categories, with equal probability.
We see that the average number of website categories in our actual data is lower than a purely probabilistic choice, for every possible publisher size.
This actively demonstrates a preferential administration of websites when it comes to their category.
Thus, publishers tend to monetize websites with similar type of content.
To verify this hypothesis, we also utilize Shannon's diversity index~\cite{shannon1948mathematical}.
Shannon's diversity index is a statistical measure, which provides information about the composition of a community.
It is defined as ${H}' = -\sum_{i=1}^{S} p_i \ln p_i$, where $S$ is the number of different categories in the dataset (\ie/ richness) and $p_i$ is the proportion of websites belonging to category $i$.
The maximum value of the diversity index is $ln(R)$, where $R$ stands for the number of distinct website categories in a community.
This represents the case where all categories are equally common inside a cluster of websites operated by the same publisher.
As we can see in Figure~\ref{fig:shannonIndex}, Shannon's diversity index for the websites in our data is much smaller than the maximum value and is closer to zero.
A smaller diversity index corresponds to a more unequal composition of the community.
We conclude that there is indeed a preferential administration or monetization and that publishers tend to acquire new websites of the same type as the ones they already manage. 
\section{Historical Analysis}
\label{sec:history}

\subsection{Historical Presence of Identifiers}
\label{sec:historical-id-presence}

We perform a historical analysis of the last two years (April 2019 to April 2021), on a trimester basis, resulting in 9 snapshots. 
We use the entire dataset of HTTPArchive~\cite{httpArchive} and do not limit ourselves to the websites found in the Tranco list. 
In Section~\ref{sec:detecting-ids}, we show that HTTP requests are a reliable source to find identifiers of interest.
Thus, we examine HTTP(S) requests of websites in these snapshots and detect \publisherIDs and \trackingIDs embedded in them. 
Figure~\ref{fig:httpArchiveDataset} illustrates our findings for the total number of websites crawled, per trimester snapshot. 
We find that, on average, 9.9\% of websites monetize their content using Google AdSense \publisherIDs, while around 64\% use Google Analytics in order to track and measure their traffic.
These trends are stable over both the snapshots and the sample size, with a standard deviation of only 0.7 and 1.81, respectively. 
These results, computed on $8\times$ more websites than the \trancoCrawl (up to $\sim$8M websites in 2021), are inline with our earlier findings described in Section~\ref{sec:methodology}, and lend credence to our analysis as being representative of the general Web. 

\begin{figure*}[t]
    \begin{minipage}[t]{0.325\textwidth}
    \centering
        \includegraphics[width=.97\textwidth]{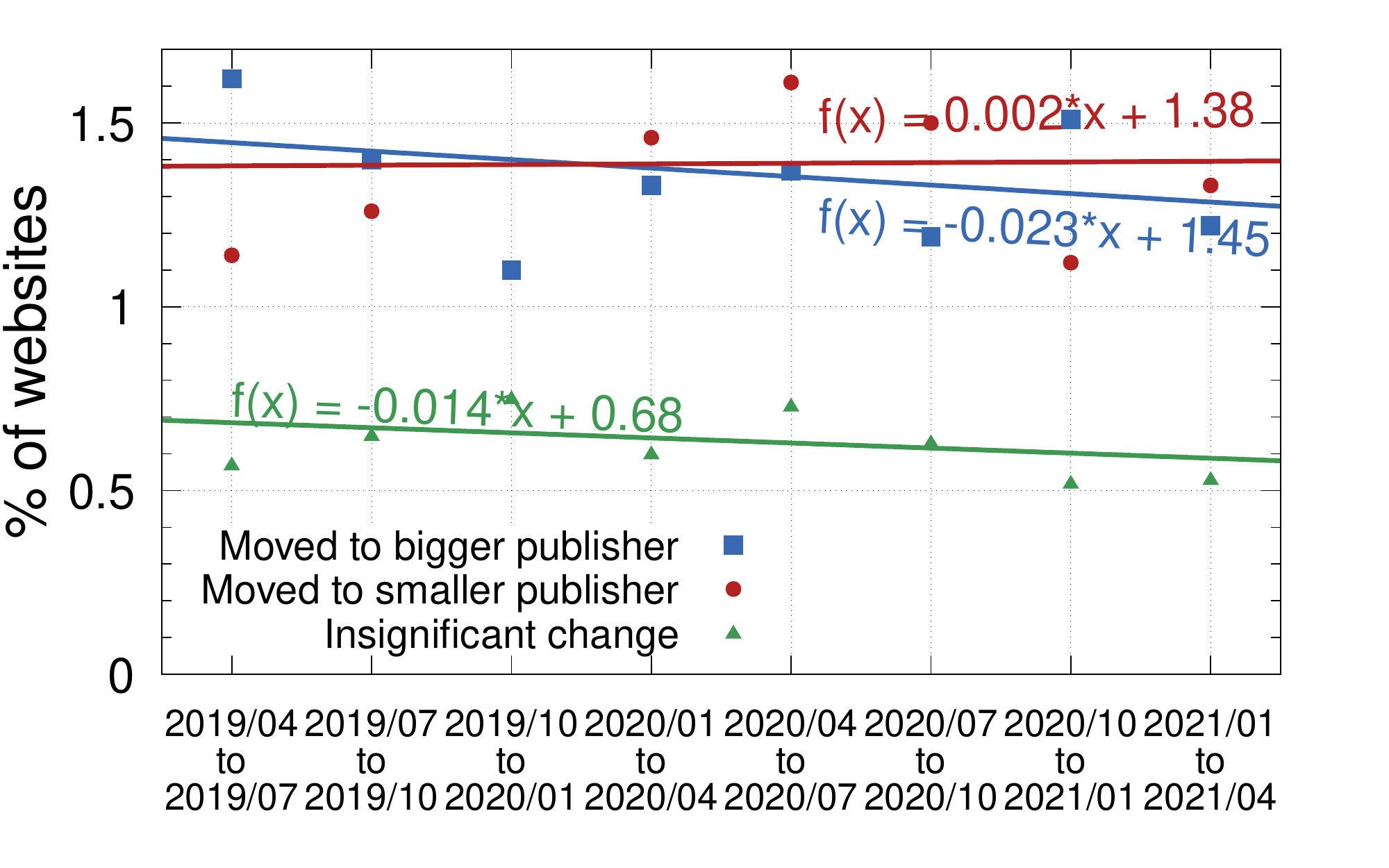}
        \caption{Transition linear trends between different size publishers through time.}
        \label{fig:historyMovementTrend}
        \end{minipage}
    \hfill
    \begin{minipage}[t]{0.325\textwidth}
    \centering
        \includegraphics[width=.97\textwidth]{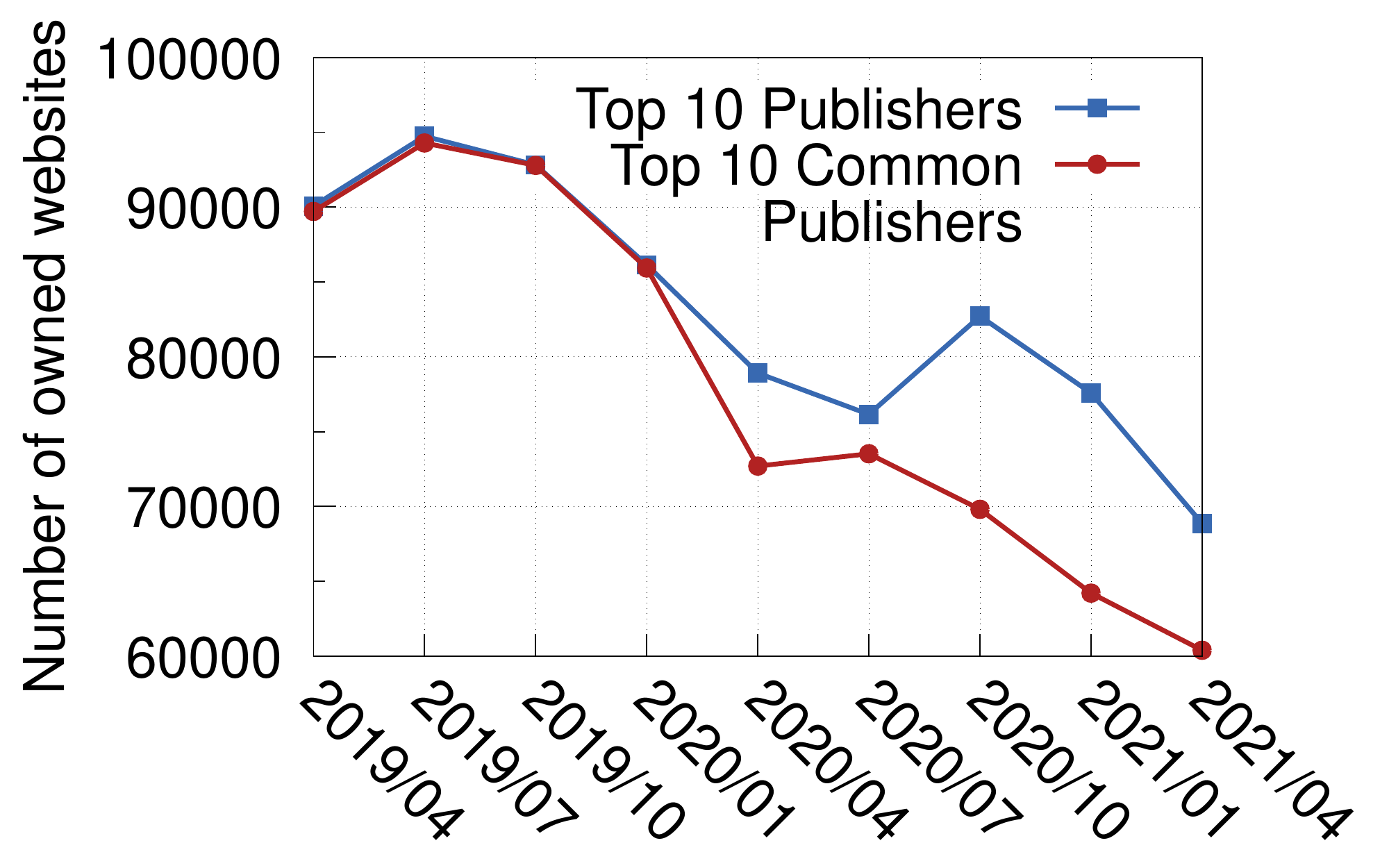}
        \caption{Number of websites monetized by the top 10 publishers in each snapshot.}
        \label{fig:historyTopPublishers}
        \end{minipage}
    \hfill
    \begin{minipage}[t]{0.325\textwidth}
    \centering
        \includegraphics[width=0.94\textwidth]{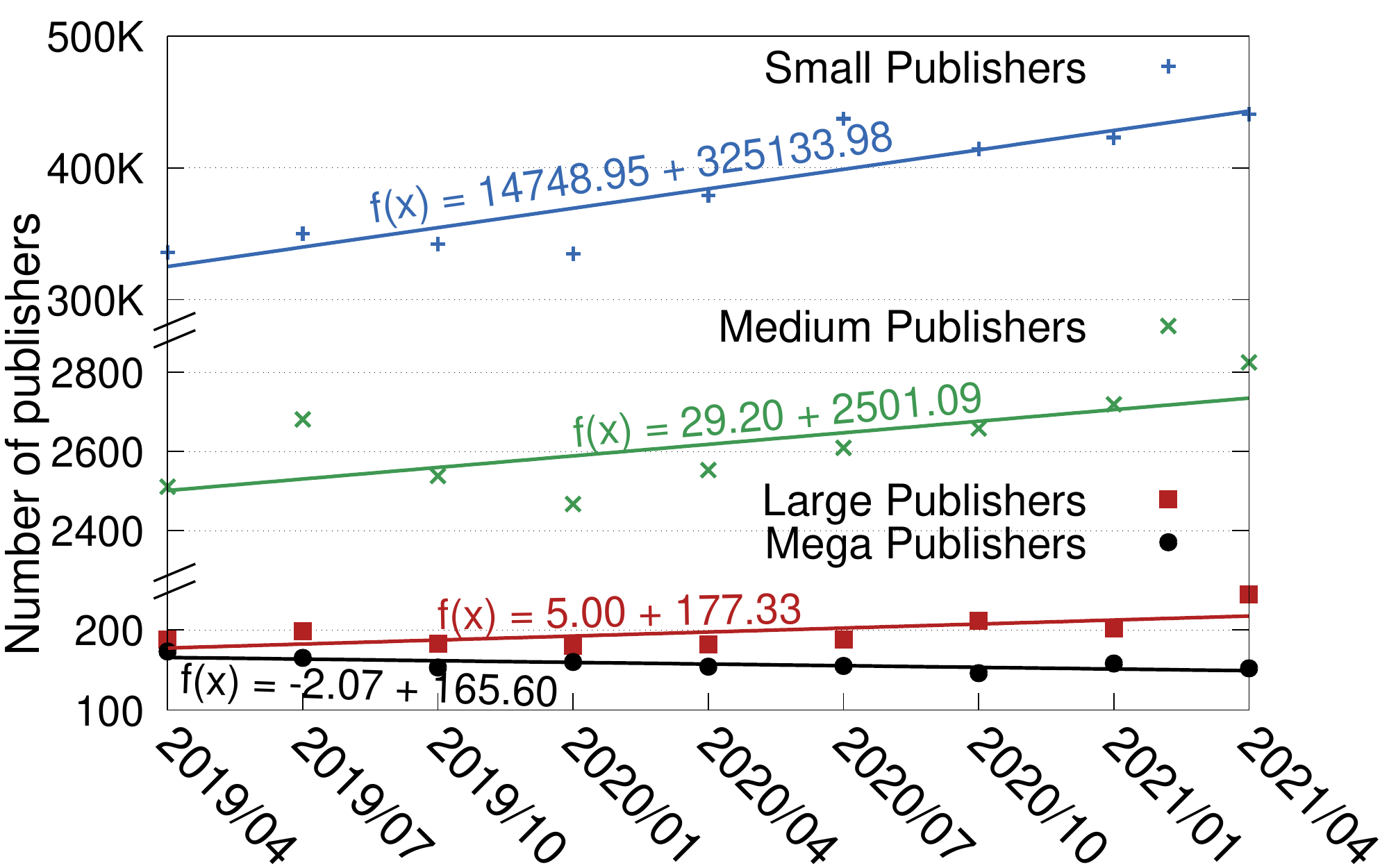}
        \caption{Changes in the population of different sizes of publishers.}
        \label{fig:historyPublisherSizes}
    \end{minipage}
    \vspace{-0.2cm}
\end{figure*}
Next, we study how many publishers contribute to the content of a website.
In Figure~\ref{fig:historyContainedIDs}, we plot the portion of websites that contain one, two, or three or more \publisherIDs for each time period. 
On the $y_2$ axis, we plot the average number of distinct \publisherIDs in each website and observe that it is almost constant across time, with a mean value of 1.11, and a standard deviation of only 0.015. 
We also find that, on average, 88.75\% of the websites, have a single contributing publisher that generates revenue. 
Finally, we find a very small amount of websites (less than 1\% in all snapshots) that contain identifiers of 3 or more publishers. 
These cases are either due to an intermediary publishing partner, or websites running under the partnership of various authors or authorized external authors.
Overall, these numbers match our earlier in-depth analysis using the \trancoCrawl of April 2021.

\subsection{Top Publishers Market}
\label{sec:historical-top-pubs}

Next, we explore how the market of publishers has changed in the last couple of years, and specifically, how intermediary publishing partners have grown.
First, we study how websites behave with regards to their \publisherIDs and detect changes of these identifiers. 
We perform our analysis on websites which have been crawled in all snapshots (\ie/ their intersection) and contain at least one \publisherID.
There are over 191,000 such websites. 
For each time interval and for each website, we compare the detected identifiers in the previous and the next snapshot. 
We ignore websites that made no change in their \publisherIDs.
For websites that do not contain exactly the same identifiers across two consecutive snapshots, we compare the size of their publishers. 
Specifically, we define the old community size of a website as the maximum size of its publishers, detected in the first snapshot for that website. 
Respectively, we retrieve the new community size from the second snapshot.
Please note that the size of a publisher for a specific snapshot is computed across all websites in the snapshot and not only for the common websites. 
If the new community size of a website is greater than the old size, then we conclude that the website moved to a bigger publisher. 
If the old size is greater, the website moved to a smaller publisher, while if the size is the same, then the publisher made an insignificant change. 
This change might be the addition or removal of a secondary contributing author, the move to a different AdSense account, \etc/
The results of this analysis can be seen in Figure~\ref{fig:historyMovement}.
We observe that the majority of websites made no changes in their \publisherIDs.
Indeed, over 96\% of websites have a consistent and stable behavior and do not change their monetization scheme. 
In contrast, we find that, on average, 3.35\% of websites made a change in their contained \publisherIDs. 

In Figure~\ref{fig:historyMovementTrend}, we show a linear regression model for the different cases of Figure~\ref{fig:historyMovement}. 
Interestingly, we find that both the cases where a website moved to a bigger publisher, and where a website made an insignificant change in their \publisherIDs, have a negative slope. 
In contrast, the case where websites move to a smaller publisher is the only one with a positive slope. 
This suggests that there is a tendency for decentralization, meaning that websites are inclined to move away from big intermediary publishers. 
To test this hypothesis, we plot in Figure~\ref{fig:historyTopPublishers} the sum of websites operated by the 10 most popular publishers in each snapshot along with the total number of websites operated by the top 10 publishers, present in all snapshots. 
We can see that there is a constant decrease in the number of websites that these ``mega-publishers'' manage. 
Indeed, in just a 2-year span, ``mega-publishers'' lost approximately 25\% of their population. 
Interestingly, this decrease in managed websites is observed even though the number of crawled websites has increased over the years (as shown in Figure~\ref{fig:httpArchiveDataset}). 

Next, we explore how this market of big publishers has changed over time.
We characterize as \textit{Small} publishers with up to 10 websites, \textit{Medium} those that monetize from 11 to 50 websites, \textit{Large} those who monetize from 51 to 100, and as \textit{Mega}, the publishers that monetize more than 100 websites. 
In Figure~\ref{fig:historyPublisherSizes}, we plot the population of these classes, \ie/ the number of such publishers, and we also fit the data subsections with a straight line. 
As we can see, the number of \textit{Small} publishers has greatly increased over the years ($\sim$15K new Small publishers per trimester), which is expected with the increasing ad-revenues motivating new independent publishers to monetize their content.
We also observe that there is an increase in the number of \textit{Medium} and \textit{Large} publishers ($\sim$29 new Medium and $\sim$5 new Large publishers per trimester), while \textit{Mega} publishers are the only class that shrinks over time ($\sim$2 Mega publishers lost per trimester). 
This is evident in the negative slope of the fitted straight line. 
This behavior attests to the fact that the market of intermediary publishing partners seems to be flourishing and that new such services have emerged during the last couple of years.
These services provide a new platform for independent content creators to generate revenue and lure clients away from \textit{Mega} publishers. 
It is evident that these new services seek their share of a competitive, but highly profitable market.
\section{Detecting Website Ownership}
\label{sec:caseStudies}

During our manually analysis, we found that there were a lot of communities which were not only operated but also owned by the same legal entity.
To better understand the utility of the \metagraph in identifying websites owned by the same legal entity, we manually examine detected communities.Overall, we find communities belonging to the news and media sector, music and entertainment, as well as manufacturing and other industrial applications.
As an example, we detect a community of websites owned by Koninklijke Philips N.V.
We find 45 official websites, each with a different country code top-level domain (ccTLD), all of which belong to the same company but serve clients of different countries.
Another community in our dataset is a cluster of 73 news websites, all serving news content using \texttt{.au} as their ccTLD.
In their privacy policy, these websites mention that they are published by a subsidiary or related body corporate of \emph{Rural Press Pty, Ltd} and, in their footer, they declare that they are operated by \emph{Australian Community Media \& Printing}.
Australian Community Media is a media company operating over 160 regional publications and targets a vast audience in multiple geographic locations.

One of the biggest detected communities is related to the music industry (\ie/ over 140 websites of popular singers or music bands).
To our surprise, these websites are owned by various companies including Atlantic Records, Electra Records, Warner Records and Nonesuch Records.
By observing the copyright notification and the privacy notices of these websites, we find that all of them are subsidiaries of a single multinational conglomerate, \emph{Warner Music Group}~\cite{warner}.
This is a clear proof that our methodology is even able to overcome the barriers of business organization and subsidiaries, and detect ownership in the highest level of hierarchy.
Finally, we find two communities of websites related to public entertainment and information.
We detect a community of 111 radio websites, operated by \emph{Townsquare Media, Inc}, a US-based radio network and media company that owns hundreds of local terrestrial radio stations~\cite{townsquare}.
We also detect a community of 76 websites, which explicitly state in their copyrights claim that they are owned by \emph{Gray Television, Inc.}, an American television broadcasting company.

These examples and many more not analyzed here due to space, demonstrate the efficacy of our methodology to detect co-ownership status of websites by organizations monetizing on them in a collective fashion.
Overall we find and report 112 distinct communities of various sizes, consisting of over 1,280 websites.
For each community we report its size, the websites that compose the community, as well as, the legal entity that owns the respective websites.
We manually visited, evaluated and verified all of these websites and make our results publicly available~\cite{openSourceData}.
We report some of the largest communities along with their size (\ie/ number of websites) and some indicative websites as examples in Appendix~\ref{sec:communitiesTable}.
\section{Related Work}
\label{sec:related}

The ecosystem of digital advertising and analytics has motivated a lot of studies that aim to reverse engineer it (\eg/~\cite{papadopoulos2017if, papadopoulos2018cost, carrascosa2015always, englehardt2016online}). 
In~\cite{gill2013best}, the authors studied the advertising ecosystem and services provided by Google, including AdSense and focused on how revenues are generated across aggregators.
In~\cite{matic2015caronte}, the authors presented an automated tool to de-anonymize Tor hidden services using information like Google Analytics and AdSense IDs to disclose the server's IP. 
Their analysis is limited regarding \identifiers, since they only extract 24 unique \analyticsIDs and 3 \publisherIDs. 
Similarly, in~\cite{yoon2019doppelgangers}, Yoon~\etal/ studied phishing threats in the DarkWeb, by trying to obtain the identity of owners operating such websites.
Using the technique of~\cite{matic2015caronte}, they extracted 276 \analyticsIDs and 1,171 \publisherIDs. 
In~\cite{starov2018betrayed}, Starov~\etal/ analyzed identifiers of multiple analytics services to bundle websites and discover malicious websites and campaigns. 
With a focus on malicious content, they identified 7,945 \analyticsIDs and 278 \containerIDs and, contrary to our work, they did not consider \publisherIDs or \measurementIDs.
In~\cite{rogers202133}, authors outlined how Google Analytics IDs can be used for digital forensics investigations to unmask online actors and lead to the entity, that operates a cluster of websites, which can be an individual, organization, or media group.

In~\cite{simeonovski2017controls}, the authors associate organizations with domain names in an attempt to create a property graph for Internet infrastructures.
To achieve this, they utilize X.509 certificates and extract the organization to which the certificate was issued. 
In~\cite{cangialosi2016measurement} authors, argue that relying on such certificates is not effective and propose a methodology that revolves around the email addresses found in WHOIS records. 
Similar to our work, authors build a bipartite graph and apply a community detection algorithm to extract clusters of domains, owned by the same organization. 
Limitations of the proposed methodology include that many WHOIS records contain the email address of the registrar or the hosting provider instead (\ie/ WHOIS privacy service).
Finally, our methodology provides an additional advantage in the cases where websites are purchased by a new legal entity, since new website owners or administrators have an incentive to update the \identifiers in their new websites in order to gain revenue or insight.
This does not apply to WHOIS records.

In~\cite{bashir2019longitudinal}, Bashir~\etal/ performed a study of the specification and adoption of \adstxt files during a 15-month period and clustered publishers serving identical \adstxt files.
Similar to our work, they found that there is a big number of smaller clusters (\ie/ less than 5 websites) but only a few big clusters with over 50 websites. 
Finally, authors manually investigated the top clusters in their dataset and found that such clusters exist due to (i) shared media properties with a common owner, (ii) independent publishers, (iii) the use of the same platform to deliver their content, or (iv) the use of consolidated SSP services.
\section{Discussion}
\label{sec:discussion}

\subsection{Summary}
In this work, we shed light on website administration by using bipartite graphs and exploiting the \identifiers that publishers embed in their websites to use third-party ad-related services.
We studied various properties induced by these graphs, reflecting important characteristics of administration, such as portfolio size, popularity, \etc/, and we identified power-law patterns of website administration, as well as indications of preferential monetization in the type of controlled websites.
We studied the use of such \identifiers across time and showed how the market of intermediary publishing partners has boomed in the last few years.
We showed that our methodology can be used to detect ownership in the Web and even overcome the company organization barriers (\ie/ subsidiary companies).

\subsection{Limitations}
Our methodology is based on detecting \identifiers using regular expressions.
However, there are cases where alphanumeric values might match with these regular expressions without being actual identifiers.
While we perform various techniques to limit these false positives (Section \ref{sec:detecting-ids}), we acknowledge that there might be cases that we miss.
Additionally, our study focuses on \identifiers related to services offered by Google, one of the biggest players in the advertising and analytics ecosystem.
Even though the analysis of Google services provides a good coverage of the real world, there are several other ad networks and analytics services that can be studied.
Finally, we acknowledge that our analysis of website categories (Section~\ref{sec:publisherid-categories}) relies on SimilarWeb, which might be prone to errors or subjective bias.

\subsection{Implications}
We believe our graph methodology and analysis is a powerful tool for web and privacy measurements aiming to understand the context, nature and activity of websites, as well as the possible leverage or political agendas behind their administration.
In fact, our proposed technique can help researchers, journalists, and even individual users to better understand popular websites and the entities that control and monetize them.
Furthermore, our preliminary analysis shows that outlier websites in the bipartite graphs yielded by our method may reveal anomalous or even malicious behavior, suggesting that our methodology can be used to discover malicious actors without even examining their published content.
Also, ad networks can make use of our technique to detect fraudulent or fake news-related website administrators that may violate their ad campaign policies.
Altogether, we believe that our method can help improve the safety and health of the Web ecosystem at large.

\section*{Acknowledgements}
This project received funding from the EU H2020 Research and Innovation programme under grant agreements No 830927 (Concordia), No 830929 (CyberSec4Europe), No 871370 (Pimcity) and No 871793 (Accordion).
These  results reflect only the authors' view and the Commission is not responsible for any use that may be made of the information it contains.

\bibliographystyle{ACM-Reference-Format}
\balance
\bibliography{main}


\begin{thebibliography}{50}


\ifx \showCODEN    \undefined \def \showCODEN     #1{\unskip}     \fi
\ifx \showDOI      \undefined \def \showDOI       #1{#1}\fi
\ifx \showISBNx    \undefined \def \showISBNx     #1{\unskip}     \fi
\ifx \showISBNxiii \undefined \def \showISBNxiii  #1{\unskip}     \fi
\ifx \showISSN     \undefined \def \showISSN      #1{\unskip}     \fi
\ifx \showLCCN     \undefined \def \showLCCN      #1{\unskip}     \fi
\ifx \shownote     \undefined \def \shownote      #1{#1}          \fi
\ifx \showarticletitle \undefined \def \showarticletitle #1{#1}   \fi
\ifx \showURL      \undefined \def \showURL       {\relax}        \fi
\providecommand\bibfield[2]{#2}
\providecommand\bibinfo[2]{#2}
\providecommand\natexlab[1]{#1}
\providecommand\showeprint[2][]{arXiv:#2}

\bibitem[\protect\citeauthoryear{Alexander}{Alexander}{2015}]%
        {globalVoices}
\bibfield{author}{\bibinfo{person}{Lawrence Alexander}.}
  \bibinfo{year}{2015}\natexlab{}.
\newblock \bibinfo{title}{Open-Source Information Reveals Pro-Kremlin Web
  Campaign}.
\newblock
  \bibinfo{howpublished}{\url{https://globalvoices.org/2015/07/13/open-source-information-reveals-pro-kremlin-web-campaign/}}.
\newblock


\bibitem[\protect\citeauthoryear{Archive}{Archive}{2021}]%
        {httpArchive}
\bibfield{author}{\bibinfo{person}{Internet Archive}.}
  \bibinfo{year}{2021}\natexlab{}.
\newblock \bibinfo{title}{HTTPArchive}.
\newblock \bibinfo{howpublished}{\url{https://httparchive.org/}}.
\newblock


\bibitem[\protect\citeauthoryear{Baio}{Baio}{2011}]%
        {wired}
\bibfield{author}{\bibinfo{person}{Andy Baio}.}
  \bibinfo{year}{2011}\natexlab{}.
\newblock \bibinfo{title}{Think You Can Hide, Anonymous Blogger? Two Words:
  Google Analytics}.
\newblock
  \bibinfo{howpublished}{\url{https://www.wired.com/2011/11/goog-analytics-anony-bloggers/}}.
\newblock


\bibitem[\protect\citeauthoryear{Bashir, Arshad, Kirda, Robertson, and
  Wilson}{Bashir et~al\mbox{.}}{2019}]%
        {bashir2019longitudinal}
\bibfield{author}{\bibinfo{person}{Muhammad~Ahmad Bashir},
  \bibinfo{person}{Sajjad Arshad}, \bibinfo{person}{Engin Kirda},
  \bibinfo{person}{William Robertson}, {and} \bibinfo{person}{Christo Wilson}.}
  \bibinfo{year}{2019}\natexlab{}.
\newblock \showarticletitle{A Longitudinal Analysis of the Ads.Txt Standard}.
  In \bibinfo{booktitle}{\emph{Proceedings of the Internet Measurement
  Conference}} (Amsterdam, Netherlands) \emph{(\bibinfo{series}{IMC '19})}.
  \bibinfo{publisher}{Association for Computing Machinery},
  \bibinfo{address}{New York, NY, USA}, \bibinfo{pages}{294–307}.
\newblock
\showISBNx{9781450369480}
\urldef\tempurl%
\url{https://doi.org/10.1145/3355369.3355603}
\showDOI{\tempurl}


\bibitem[\protect\citeauthoryear{Blondel, Guillaume, Lambiotte, and
  Lefebvre}{Blondel et~al\mbox{.}}{2008}]%
        {blondel2008fast}
\bibfield{author}{\bibinfo{person}{Vincent~D Blondel},
  \bibinfo{person}{Jean-Loup Guillaume}, \bibinfo{person}{Renaud Lambiotte},
  {and} \bibinfo{person}{Etienne Lefebvre}.} \bibinfo{year}{2008}\natexlab{}.
\newblock \showarticletitle{Fast unfolding of communities in large networks}.
\newblock \bibinfo{journal}{\emph{Journal of statistical mechanics: theory and
  experiment}} \bibinfo{volume}{2008}, \bibinfo{number}{10}
  (\bibinfo{year}{2008}), \bibinfo{pages}{P10008}.
\newblock


\bibitem[\protect\citeauthoryear{Cangialosi, Chung, Choffnes, Levin, Maggs,
  Mislove, and Wilson}{Cangialosi et~al\mbox{.}}{2016}]%
        {cangialosi2016measurement}
\bibfield{author}{\bibinfo{person}{Frank Cangialosi}, \bibinfo{person}{Taejoong
  Chung}, \bibinfo{person}{David Choffnes}, \bibinfo{person}{Dave Levin},
  \bibinfo{person}{Bruce~M. Maggs}, \bibinfo{person}{Alan Mislove}, {and}
  \bibinfo{person}{Christo Wilson}.} \bibinfo{year}{2016}\natexlab{}.
\newblock \showarticletitle{Measurement and Analysis of Private Key Sharing in
  the HTTPS Ecosystem}. In \bibinfo{booktitle}{\emph{Proceedings of the 2016
  ACM SIGSAC Conference on Computer and Communications Security}} (Vienna,
  Austria) \emph{(\bibinfo{series}{CCS '16})}. \bibinfo{publisher}{Association
  for Computing Machinery}, \bibinfo{address}{New York, NY, USA},
  \bibinfo{pages}{628–640}.
\newblock
\showISBNx{9781450341394}
\urldef\tempurl%
\url{https://doi.org/10.1145/2976749.2978301}
\showDOI{\tempurl}


\bibitem[\protect\citeauthoryear{Carrascosa, Mikians, Cuevas, Erramilli, and
  Laoutaris}{Carrascosa et~al\mbox{.}}{2015}]%
        {carrascosa2015always}
\bibfield{author}{\bibinfo{person}{Juan~Miguel Carrascosa},
  \bibinfo{person}{Jakub Mikians}, \bibinfo{person}{Ruben Cuevas},
  \bibinfo{person}{Vijay Erramilli}, {and} \bibinfo{person}{Nikolaos
  Laoutaris}.} \bibinfo{year}{2015}\natexlab{}.
\newblock \showarticletitle{I Always Feel like Somebody's Watching Me:
  Measuring Online Behavioural Advertising}. In
  \bibinfo{booktitle}{\emph{Proceedings of the 11th ACM Conference on Emerging
  Networking Experiments and Technologies}} (Heidelberg, Germany)
  \emph{(\bibinfo{series}{CoNEXT '15})}. \bibinfo{publisher}{Association for
  Computing Machinery}, \bibinfo{address}{New York, NY, USA}, Article
  \bibinfo{articleno}{13}, \bibinfo{numpages}{13}~pages.
\newblock
\showISBNx{9781450334129}
\urldef\tempurl%
\url{https://doi.org/10.1145/2716281.2836098}
\showDOI{\tempurl}


\bibitem[\protect\citeauthoryear{Center}{Center}{2021a}]%
        {tags}
\bibfield{author}{\bibinfo{person}{Google~Help Center}.}
  \bibinfo{year}{2021}\natexlab{a}.
\newblock \bibinfo{title}{About Tags}.
\newblock
  \bibinfo{howpublished}{\url{https://support.google.com/tagmanager/answer/3281060}}.
\newblock


\bibitem[\protect\citeauthoryear{Center}{Center}{2021b}]%
        {adsenseFaq}
\bibfield{author}{\bibinfo{person}{Google~Help Center}.}
  \bibinfo{year}{2021}\natexlab{b}.
\newblock \bibinfo{title}{Ad placement policies}.
\newblock
  \bibinfo{howpublished}{\url{https://support.google.com/adsense/answer/2659106}}.
\newblock


\bibitem[\protect\citeauthoryear{Center}{Center}{2021c}]%
        {adsenseAccess}
\bibfield{author}{\bibinfo{person}{Google~Help Center}.}
  \bibinfo{year}{2021}\natexlab{c}.
\newblock \bibinfo{title}{Manage user access to your account}.
\newblock
  \bibinfo{howpublished}{\url{https://support.google.com/adsense/answer/2646544}}.
\newblock


\bibitem[\protect\citeauthoryear{Center}{Center}{2021d}]%
        {containers}
\bibfield{author}{\bibinfo{person}{Google~Help Center}.}
  \bibinfo{year}{2021}\natexlab{d}.
\newblock \bibinfo{title}{Organize your containers}.
\newblock
  \bibinfo{howpublished}{\url{https://support.google.com/tagmanager/answer/6261285}}.
\newblock


\bibitem[\protect\citeauthoryear{Center}{Center}{2021e}]%
        {adsenseShare}
\bibfield{author}{\bibinfo{person}{Google~Help Center}.}
  \bibinfo{year}{2021}\natexlab{e}.
\newblock \bibinfo{title}{Revenue share}.
\newblock
  \bibinfo{howpublished}{\url{https://support.google.com/adsense/answer/1346295}}.
\newblock


\bibitem[\protect\citeauthoryear{Center}{Center}{2021f}]%
        {gtm}
\bibfield{author}{\bibinfo{person}{Google~Help Center}.}
  \bibinfo{year}{2021}\natexlab{f}.
\newblock \bibinfo{title}{Setup and install Tag Manager}.
\newblock
  \bibinfo{howpublished}{\url{https://support.google.com/tagmanager/answer/6103696}}.
\newblock


\bibitem[\protect\citeauthoryear{Center}{Center}{2021g}]%
        {trackingID}
\bibfield{author}{\bibinfo{person}{Google~Help Center}.}
  \bibinfo{year}{2021}\natexlab{g}.
\newblock \bibinfo{title}{Tracking ID and property number}.
\newblock
  \bibinfo{howpublished}{\url{https://support.google.com/analytics/answer/7372977}}.
\newblock


\bibitem[\protect\citeauthoryear{Center}{Center}{2021h}]%
        {universalAnalytics}
\bibfield{author}{\bibinfo{person}{Google~Help Center}.}
  \bibinfo{year}{2021}\natexlab{h}.
\newblock \bibinfo{title}{Universal Analytics property}.
\newblock
  \bibinfo{howpublished}{\url{https://support.google.com/analytics/answer/10220206}}.
\newblock


\bibitem[\protect\citeauthoryear{Clauset, Shalizi, and Newman}{Clauset
  et~al\mbox{.}}{2009}]%
        {clauset2009power}
\bibfield{author}{\bibinfo{person}{Aaron Clauset},
  \bibinfo{person}{Cosma~Rohilla Shalizi}, {and} \bibinfo{person}{Mark~EJ
  Newman}.} \bibinfo{year}{2009}\natexlab{}.
\newblock \showarticletitle{Power-law distributions in empirical data}.
\newblock \bibinfo{journal}{\emph{SIAM review}} \bibinfo{volume}{51},
  \bibinfo{number}{4} (\bibinfo{year}{2009}), \bibinfo{pages}{661--703}.
\newblock


\bibitem[\protect\citeauthoryear{Community}{Community}{2019}]%
        {bloggerAds}
\bibfield{author}{\bibinfo{person}{AdSense~Help Community}.}
  \bibinfo{year}{2019}\natexlab{}.
\newblock \bibinfo{title}{What is the adsenseHostId on blogger?}
\newblock
  \bibinfo{howpublished}{\url{https://support.google.com/adsense/thread/18637422/what-is-the-adsensehostid-on-blogger-why-there-is-another-pub-id}}.
\newblock


\bibitem[\protect\citeauthoryear{Cramer-Flood}{Cramer-Flood}{2021}]%
        {adSpending}
\bibfield{author}{\bibinfo{person}{Ethan Cramer-Flood}.}
  \bibinfo{year}{2021}\natexlab{}.
\newblock \bibinfo{title}{Worldwide Digital Ad Spending 2021}.
\newblock
  \bibinfo{howpublished}{\url{https://www.emarketer.com/content/worldwide-digital-ad-spending-2021}}.
\newblock


\bibitem[\protect\citeauthoryear{Dittrich and Kenneally}{Dittrich and
  Kenneally}{2012}]%
        {dittrich2012menloreport}
\bibfield{author}{\bibinfo{person}{David Dittrich} {and} \bibinfo{person}{Erin
  Kenneally}.} \bibinfo{year}{2012}\natexlab{}.
\newblock \bibinfo{booktitle}{\emph{The Menlo Report: Ethical Principles
  Guiding Information and Communication Technology Research}}.
\newblock \bibinfo{type}{{T}echnical {R}eport}.
\newblock


\bibitem[\protect\citeauthoryear{Englehardt and Narayanan}{Englehardt and
  Narayanan}{2016}]%
        {englehardt2016online}
\bibfield{author}{\bibinfo{person}{Steven Englehardt} {and}
  \bibinfo{person}{Arvind Narayanan}.} \bibinfo{year}{2016}\natexlab{}.
\newblock \showarticletitle{Online Tracking: A 1-Million-Site Measurement and
  Analysis}. In \bibinfo{booktitle}{\emph{Proceedings of the 2016 ACM SIGSAC
  Conference on Computer and Communications Security}} (Vienna, Austria)
  \emph{(\bibinfo{series}{CCS '16})}. \bibinfo{publisher}{Association for
  Computing Machinery}, \bibinfo{address}{New York, NY, USA},
  \bibinfo{pages}{1388–1401}.
\newblock
\showISBNx{9781450341394}
\urldef\tempurl%
\url{https://doi.org/10.1145/2976749.2978313}
\showDOI{\tempurl}


\bibitem[\protect\citeauthoryear{Farney}{Farney}{2016}]%
        {farney2016google}
\bibfield{author}{\bibinfo{person}{Tabatha Farney}.}
  \bibinfo{year}{2016}\natexlab{}.
\newblock \bibinfo{booktitle}{\emph{Google Analytics and Google Tag Manager}}.
\newblock \bibinfo{publisher}{ALA TechSource}, \bibinfo{address}{Chicago}.
\newblock
\showISBNx{978-0-8389-5976-3}


\bibitem[\protect\citeauthoryear{Gill, Erramilli, Chaintreau, Krishnamurthy,
  Papagiannaki, and Rodriguez}{Gill et~al\mbox{.}}{2013}]%
        {gill2013best}
\bibfield{author}{\bibinfo{person}{Phillipa Gill}, \bibinfo{person}{Vijay
  Erramilli}, \bibinfo{person}{Augustin Chaintreau},
  \bibinfo{person}{Balachander Krishnamurthy}, \bibinfo{person}{Konstantina
  Papagiannaki}, {and} \bibinfo{person}{Pablo Rodriguez}.}
  \bibinfo{year}{2013}\natexlab{}.
\newblock \showarticletitle{Best Paper -- Follow the Money: Understanding
  Economics of Online Aggregation and Advertising}. In
  \bibinfo{booktitle}{\emph{Proceedings of the 2013 Conference on Internet
  Measurement Conference}} (Barcelona, Spain) \emph{(\bibinfo{series}{IMC
  '13})}. \bibinfo{publisher}{Association for Computing Machinery},
  \bibinfo{address}{New York, NY, USA}, \bibinfo{pages}{141–148}.
\newblock
\showISBNx{9781450319539}
\urldef\tempurl%
\url{https://doi.org/10.1145/2504730.2504768}
\showDOI{\tempurl}


\bibitem[\protect\citeauthoryear{Girvan and Newman}{Girvan and Newman}{2002}]%
        {girvan2002community}
\bibfield{author}{\bibinfo{person}{Michelle Girvan} {and}
  \bibinfo{person}{Mark~EJ Newman}.} \bibinfo{year}{2002}\natexlab{}.
\newblock \showarticletitle{Community structure in social and biological
  networks}.
\newblock \bibinfo{journal}{\emph{Proceedings of the national academy of
  sciences}} \bibinfo{volume}{99}, \bibinfo{number}{12} (\bibinfo{year}{2002}),
  \bibinfo{pages}{7821--7826}.
\newblock


\bibitem[\protect\citeauthoryear{GoDaddy Media~Temple}{GoDaddy
  Media~Temple}{2021a}]%
        {sucuri}
\bibfield{author}{\bibinfo{person}{Inc. GoDaddy Media~Temple}.}
  \bibinfo{year}{2021}\natexlab{a}.
\newblock \bibinfo{title}{Sucuri - Free website security check and malware
  scanner}.
\newblock \bibinfo{howpublished}{\url{https://sitecheck.sucuri.net/}}.
\newblock


\bibitem[\protect\citeauthoryear{GoDaddy Media~Temple}{GoDaddy
  Media~Temple}{2021b}]%
        {sucuriReport}
\bibfield{author}{\bibinfo{person}{Inc. GoDaddy Media~Temple}.}
  \bibinfo{year}{2021}\natexlab{b}.
\newblock \bibinfo{title}{Sucuri Report - prykoly.ru}.
\newblock
  \bibinfo{howpublished}{\url{https://sitecheck.sucuri.net/results/prykoly.ru}}.
\newblock


\bibitem[\protect\citeauthoryear{Google}{Google}{2021}]%
        {googlePartners}
\bibfield{author}{\bibinfo{person}{Google}.} \bibinfo{year}{2021}\natexlab{}.
\newblock \bibinfo{title}{Certified Publishing Partner}.
\newblock
  \bibinfo{howpublished}{\url{https://www.google.com/ads/publisher/partners/}}.
\newblock


\bibitem[\protect\citeauthoryear{Group}{Group}{2021}]%
        {warner}
\bibfield{author}{\bibinfo{person}{Warner~Music Group}.}
  \bibinfo{year}{2021}\natexlab{}.
\newblock \bibinfo{title}{Services - Recorded Music}.
\newblock \bibinfo{howpublished}{\url{https://www.wmg.com/services}}.
\newblock


\bibitem[\protect\citeauthoryear{Hwang}{Hwang}{2020}]%
        {hwang2020adtimebomb}
\bibfield{author}{\bibinfo{person}{Tim Hwang}.}
  \bibinfo{year}{2020}\natexlab{}.
\newblock \bibinfo{booktitle}{\emph{Subprime Attention Crisis: Advertising and
  the Time Bomb at the Heart of the Internet}}.
\newblock \bibinfo{publisher}{FSG Originals x Logic, Farrar, Straus and
  Giroux}, \bibinfo{address}{New York}.
\newblock
\showISBNx{9780374538651}


\bibitem[\protect\citeauthoryear{Laboratory}{Laboratory}{2021}]%
        {iabAdRevenue-covid-2020}
\bibfield{author}{\bibinfo{person}{IAB~Technology Laboratory}.}
  \bibinfo{year}{2021}\natexlab{}.
\newblock \bibinfo{title}{IAB Releases Internet Advertising Revenue Report for
  2020}.
\newblock
  \bibinfo{howpublished}{\url{https://www.iab.com/news/iab-internet-advertising-revenue/}}.
\newblock


\bibitem[\protect\citeauthoryear{Limited}{Limited}{[n.d.]}]%
        {virustotal}
\bibfield{author}{\bibinfo{person}{Chronicle Security~Ireland Limited}.}
  \bibinfo{year}{[n.d.]}\natexlab{}.
\newblock \bibinfo{title}{VirusTotal - Analyze suspicious files and URLs to
  detect types of malware, automatically share them with the security
  community}.
\newblock \bibinfo{howpublished}{\url{https://www.virustotal.com/}}.
\newblock


\bibitem[\protect\citeauthoryear{Limited}{Limited}{2021}]%
        {virustotalReport}
\bibfield{author}{\bibinfo{person}{Chronicle Security~Ireland Limited}.}
  \bibinfo{year}{2021}\natexlab{}.
\newblock \bibinfo{title}{VirusTotal Report - pps.net}.
\newblock
  \bibinfo{howpublished}{\url{https://www.virustotal.com/gui/domain/www.pps.net}}.
\newblock


\bibitem[\protect\citeauthoryear{LTD}{LTD}{[n.d.]}]%
        {similarWeb}
\bibfield{author}{\bibinfo{person}{Similarweb LTD}.}
  \bibinfo{year}{[n.d.]}\natexlab{}.
\newblock \bibinfo{title}{Website Traffic - Check and Analyze Any Website}.
\newblock \bibinfo{howpublished}{\url{https://www.similarweb.com/}}.
\newblock


\bibitem[\protect\citeauthoryear{Matic, Kotzias, and Caballero}{Matic
  et~al\mbox{.}}{2015}]%
        {matic2015caronte}
\bibfield{author}{\bibinfo{person}{Srdjan Matic}, \bibinfo{person}{Platon
  Kotzias}, {and} \bibinfo{person}{Juan Caballero}.}
  \bibinfo{year}{2015}\natexlab{}.
\newblock \showarticletitle{CARONTE: Detecting Location Leaks for Deanonymizing
  Tor Hidden Services}. In \bibinfo{booktitle}{\emph{Proceedings of the 22nd
  ACM SIGSAC Conference on Computer and Communications Security}} (Denver,
  Colorado, USA) \emph{(\bibinfo{series}{CCS '15})}.
  \bibinfo{publisher}{Association for Computing Machinery},
  \bibinfo{address}{New York, NY, USA}, \bibinfo{pages}{1455–1466}.
\newblock
\showISBNx{9781450338325}
\urldef\tempurl%
\url{https://doi.org/10.1145/2810103.2813667}
\showDOI{\tempurl}


\bibitem[\protect\citeauthoryear{Media}{Media}{2021}]%
        {townsquare}
\bibfield{author}{\bibinfo{person}{Townsquare Media}.}
  \bibinfo{year}{2021}\natexlab{}.
\newblock \bibinfo{title}{Digital Media and Radio Advertising Company}.
\newblock \bibinfo{howpublished}{\url{https://www.townsquaremedia.com/}}.
\newblock


\bibitem[\protect\citeauthoryear{Papadogiannakis}{Papadogiannakis}{2021a}]%
        {openSourceCode}
\bibfield{author}{\bibinfo{person}{Emmanouil Papadogiannakis}.}
  \bibinfo{year}{2021}\natexlab{a}.
\newblock \bibinfo{title}{Scrape Titan}.
\newblock
  \bibinfo{howpublished}{\url{https://gitlab.com/papamano/scrape-titan}}.
\newblock


\bibitem[\protect\citeauthoryear{Papadogiannakis}{Papadogiannakis}{2021b}]%
        {openSourceData}
\bibfield{author}{\bibinfo{person}{Emmanouil Papadogiannakis}.}
  \bibinfo{year}{2021}\natexlab{b}.
\newblock \bibinfo{title}{Website Administration Graphs}.
\newblock
  \bibinfo{howpublished}{\url{https://gitlab.com/papamano/website-administration-graphs}}.
\newblock


\bibitem[\protect\citeauthoryear{Papadopoulos, Kourtellis, and
  Markatos}{Papadopoulos et~al\mbox{.}}{2018}]%
        {papadopoulos2018cost}
\bibfield{author}{\bibinfo{person}{Panagiotis Papadopoulos},
  \bibinfo{person}{Nicolas Kourtellis}, {and} \bibinfo{person}{Evangelos~P.
  Markatos}.} \bibinfo{year}{2018}\natexlab{}.
\newblock \showarticletitle{The Cost of Digital Advertisement: Comparing User
  and Advertiser Views}. In \bibinfo{booktitle}{\emph{Proceedings of the 2018
  World Wide Web Conference}} (Lyon, France) \emph{(\bibinfo{series}{WWW
  '18})}. \bibinfo{publisher}{International World Wide Web Conferences Steering
  Committee}, \bibinfo{address}{Republic and Canton of Geneva, CHE},
  \bibinfo{pages}{1479–1489}.
\newblock
\showISBNx{9781450356398}
\urldef\tempurl%
\url{https://doi.org/10.1145/3178876.3186060}
\showDOI{\tempurl}


\bibitem[\protect\citeauthoryear{Papadopoulos, Kourtellis, Rodriguez, and
  Laoutaris}{Papadopoulos et~al\mbox{.}}{2017}]%
        {papadopoulos2017if}
\bibfield{author}{\bibinfo{person}{Panagiotis Papadopoulos},
  \bibinfo{person}{Nicolas Kourtellis}, \bibinfo{person}{Pablo~Rodriguez
  Rodriguez}, {and} \bibinfo{person}{Nikolaos Laoutaris}.}
  \bibinfo{year}{2017}\natexlab{}.
\newblock \showarticletitle{If You Are Not Paying for It, You Are the Product:
  How Much Do Advertisers Pay to Reach You?}. In
  \bibinfo{booktitle}{\emph{Proceedings of the 2017 Internet Measurement
  Conference}} (London, United Kingdom) \emph{(\bibinfo{series}{IMC '17})}.
  \bibinfo{publisher}{Association for Computing Machinery},
  \bibinfo{address}{New York, NY, USA}, \bibinfo{pages}{142–156}.
\newblock
\showISBNx{9781450351188}
\urldef\tempurl%
\url{https://doi.org/10.1145/3131365.3131397}
\showDOI{\tempurl}


\bibitem[\protect\citeauthoryear{Project and Atkinson}{Project and
  Atkinson}{2019}]%
        {aspell}
\bibfield{author}{\bibinfo{person}{GNU Project} {and} \bibinfo{person}{Kevin
  Atkinson}.} \bibinfo{year}{2019}\natexlab{}.
\newblock \bibinfo{title}{GNU Aspell - Free and Open Source spell checker}.
\newblock \bibinfo{howpublished}{\url{http://aspell.net/}}.
\newblock


\bibitem[\protect\citeauthoryear{PublicWWW}{PublicWWW}{[n.d.]}]%
        {publicWWW}
\bibfield{author}{\bibinfo{person}{PublicWWW}.}
  \bibinfo{year}{[n.d.]}\natexlab{}.
\newblock \bibinfo{title}{Source Code Search Engine}.
\newblock \bibinfo{howpublished}{\url{https://publicwww.com/}}.
\newblock


\bibitem[\protect\citeauthoryear{Rivers and Lewis}{Rivers and Lewis}{2014}]%
        {rivers2014ethicalresearchstandards}
\bibfield{author}{\bibinfo{person}{Caitlin~M. Rivers} {and}
  \bibinfo{person}{Bryan~L. Lewis}.} \bibinfo{year}{2014}\natexlab{}.
\newblock \showarticletitle{Ethical research standards in a world of big data}.
\newblock \bibinfo{journal}{\emph{F1000Research}}  \bibinfo{volume}{3}
  (\bibinfo{year}{2014}), \bibinfo{pages}{38}.
\newblock
\urldef\tempurl%
\url{https://doi.org/10.12688/f1000research.3-38.v2}
\showDOI{\tempurl}


\bibitem[\protect\citeauthoryear{Rogers}{Rogers}{2021}]%
        {rogers202133}
\bibfield{author}{\bibinfo{person}{Richard Rogers}.}
  \bibinfo{year}{2021}\natexlab{}.
\newblock \bibinfo{booktitle}{\emph{Digital Forensics: Repurposing Google
  Analytics IDs}}.
\newblock \bibinfo{publisher}{Amsterdam University Press},
  \bibinfo{address}{Amsterdam,The Netherlands}, \bibinfo{pages}{241--245}.
\newblock
\showISBNx{9789462989511}
\urldef\tempurl%
\url{http://www.jstor.org/stable/j.ctv1qr6smr.36}
\showURL{%
\tempurl}


\bibitem[\protect\citeauthoryear{Samson}{Samson}{2018}]%
        {verafiles}
\bibfield{author}{\bibinfo{person}{Celine~Isabelle Samson}.}
  \bibinfo{year}{2018}\natexlab{}.
\newblock \bibinfo{title}{VERA FILES FACT CHECK YEARENDER: Ads reveal links
  between websites producing fake news}.
\newblock
  \bibinfo{howpublished}{\url{https://www.verafiles.org/articles/vera-files-fact-check-yearender-ads-reveal-links-between-web}}.
\newblock


\bibitem[\protect\citeauthoryear{Shannon}{Shannon}{1948}]%
        {shannon1948mathematical}
\bibfield{author}{\bibinfo{person}{Claude~Elwood Shannon}.}
  \bibinfo{year}{1948}\natexlab{}.
\newblock \showarticletitle{A mathematical theory of communication}.
\newblock \bibinfo{journal}{\emph{The Bell system technical journal}}
  \bibinfo{volume}{27}, \bibinfo{number}{3} (\bibinfo{year}{1948}),
  \bibinfo{pages}{379--423}.
\newblock


\bibitem[\protect\citeauthoryear{Silverman, Lytvynenko, Vo, and
  Singer-Vine}{Silverman et~al\mbox{.}}{2017}]%
        {buzzfeed}
\bibfield{author}{\bibinfo{person}{Craig Silverman}, \bibinfo{person}{Jane
  Lytvynenko}, \bibinfo{person}{Lam~Thuy Vo}, {and} \bibinfo{person}{Jeremy
  Singer-Vine}.} \bibinfo{year}{2017}\natexlab{}.
\newblock \bibinfo{title}{Inside The Partisan Fight For Your News Feed}.
\newblock
  \bibinfo{howpublished}{\url{https://www.buzzfeednews.com/article/craigsilverman/inside-the-partisan-fight-for-your-news-feed}}.
\newblock


\bibitem[\protect\citeauthoryear{Simeonovski, Pellegrino, Rossow, and
  Backes}{Simeonovski et~al\mbox{.}}{2017}]%
        {simeonovski2017controls}
\bibfield{author}{\bibinfo{person}{Milivoj Simeonovski},
  \bibinfo{person}{Giancarlo Pellegrino}, \bibinfo{person}{Christian Rossow},
  {and} \bibinfo{person}{Michael Backes}.} \bibinfo{year}{2017}\natexlab{}.
\newblock \showarticletitle{Who Controls the Internet? Analyzing Global Threats
  Using Property Graph Traversals}. In \bibinfo{booktitle}{\emph{Proceedings of
  the 26th International Conference on World Wide Web}} (Perth, Australia)
  \emph{(\bibinfo{series}{WWW '17})}. \bibinfo{publisher}{International World
  Wide Web Conferences Steering Committee}, \bibinfo{address}{Republic and
  Canton of Geneva, CHE}, \bibinfo{pages}{647–656}.
\newblock
\showISBNx{9781450349130}
\urldef\tempurl%
\url{https://doi.org/10.1145/3038912.3052587}
\showDOI{\tempurl}


\bibitem[\protect\citeauthoryear{Starov, Zhou, Zhang, Miramirkhani, and
  Nikiforakis}{Starov et~al\mbox{.}}{2018}]%
        {starov2018betrayed}
\bibfield{author}{\bibinfo{person}{Oleksii Starov}, \bibinfo{person}{Yuchen
  Zhou}, \bibinfo{person}{Xiao Zhang}, \bibinfo{person}{Najmeh Miramirkhani},
  {and} \bibinfo{person}{Nick Nikiforakis}.} \bibinfo{year}{2018}\natexlab{}.
\newblock \showarticletitle{Betrayed by Your Dashboard: Discovering Malicious
  Campaigns via Web Analytics}. In \bibinfo{booktitle}{\emph{Proceedings of the
  2018 World Wide Web Conference}} (Lyon, France) \emph{(\bibinfo{series}{WWW
  '18})}. \bibinfo{publisher}{International World Wide Web Conferences Steering
  Committee}, \bibinfo{address}{Republic and Canton of Geneva, CHE},
  \bibinfo{pages}{227–236}.
\newblock
\showISBNx{9781450356398}
\urldef\tempurl%
\url{https://doi.org/10.1145/3178876.3186089}
\showDOI{\tempurl}


\bibitem[\protect\citeauthoryear{{The Chromium Authors}}{{The Chromium
  Authors}}{2014}]%
        {chromeDevTools}
\bibfield{author}{\bibinfo{person}{{The Chromium Authors}}.}
  \bibinfo{year}{2014}\natexlab{}.
\newblock \bibinfo{title}{Chrome DevTools Protocol}.
\newblock
  \bibinfo{howpublished}{\url{https://chromedevtools.github.io/devtools-protocol/}}.
\newblock


\bibitem[\protect\citeauthoryear{{Tranco}}{{Tranco}}{2021}]%
        {trancoList}
\bibfield{author}{\bibinfo{person}{{Tranco}}.} \bibinfo{year}{2021}\natexlab{}.
\newblock \bibinfo{title}{Tranco list with the 1M top sites generated on 14
  April 2021}.
\newblock \bibinfo{howpublished}{\url{https://tranco-list.eu/list/7JVX/full}}.
\newblock


\bibitem[\protect\citeauthoryear{Yoon, Kim, Kim, Shin, and Son}{Yoon
  et~al\mbox{.}}{2019}]%
        {yoon2019doppelgangers}
\bibfield{author}{\bibinfo{person}{Changhoon Yoon}, \bibinfo{person}{Kwanwoo
  Kim}, \bibinfo{person}{Yongdae Kim}, \bibinfo{person}{Seungwon Shin}, {and}
  \bibinfo{person}{Sooel Son}.} \bibinfo{year}{2019}\natexlab{}.
\newblock \showarticletitle{Doppelg\"{a}Ngers on the Dark Web: A Large-Scale
  Assessment on Phishing Hidden Web Services}. In \bibinfo{booktitle}{\emph{The
  World Wide Web Conference}} (San Francisco, CA, USA)
  \emph{(\bibinfo{series}{WWW '19})}. \bibinfo{publisher}{Association for
  Computing Machinery}, \bibinfo{address}{New York, NY, USA},
  \bibinfo{pages}{2225–2235}.
\newblock
\showISBNx{9781450366748}
\urldef\tempurl%
\url{https://doi.org/10.1145/3308558.3313551}
\showDOI{\tempurl}


\end{thebibliography}

\appendix
\section{Ethical Considerations}
\label{sec:ethics}

The execution of this work has followed the principles and guidelines of how to perform ethical information research and the use of shared measurement data~\cite{dittrich2012menloreport,rivers2014ethicalresearchstandards}.
In particular, this study paid attention to the following dimensions.

We keep our crawling to a minimum to ensure that we do not slow down or deteriorate the performance of any web service in any way.
Therefore, we crawl only the landing page of each website and visit it only once.
We do not interact with any component in the visited website, and only passively observe network traffic.
In addition to this, our crawler has been implemented to wait for both the website to fully load and an extra period of time before visiting another website.
Consequently, we emulate the behavior of a normal user that stumbled upon a website.

In accordance to the GDPR and ePrivacy regulations, we did not engage in collection of data from real users.
Also, we do not share with any other entity any data collected by our crawler.
Our analysis is, to a large extent, based on public historical data (\eg/ HTTPArchive Project).
Moreover, we ensure that the privacy of publishers and administrators is not invaded.
We do not collect any of their information (\eg/ email addresses) and only discuss publishers, who explicitly and voluntarily disclose their identity in their websites, as we did in Section~\ref{sec:caseStudies}.
Last but not least, we intentionally do not make our \trancoCrawl dataset public, to ensure that there is no infringement of copyrighted material.

\section{Detected Communities}
\label{sec:communitiesTable}

\begin{table}[h] 
\footnotesize
\caption{Communities detected using the proposed methodology. Through manual investigation we determine the legal entity behind these websites.}
    \centering
    \begin{tabular}{lrl}
    \toprule
        \textbf{Description} & \textbf{Size} & \textbf{Websites} \\
    \midrule
        MinuteMedia & 172 & showsnob.com, sodomojo.com,\\
        && sportdfw.com, thejetpress.com,\\
        && reignoftroy.com, 90min.de, ... \\
        
        Warner Music Group  & 142 & brunomars.com, blakeshelton.com,\\
        && greenday.com, vancejoy.com,\\
        && paramore.net, disturbed1.com, ...\\
        
        Townsquare Media, Inc & 111 & wkdq.com, wgrd.com, wbkr.com,\\
        && wbckfm.com, mix108.com, b985.fm,\\
        && keyw.com, 929nin.com, ...\\
        
        Gray Media Group, Inc & 76 & wrdw.com, witn.com, whsv.com,\\
        && wcax.com, wbay.com, nbc29.com,\\
        && kwch.com, kktv.com, abc12.com,...\\
        
        Australian Community Media  & 73 & thesenior.com.au, nvi.com.au,\\
        && theflindersnews.com.au, \\
        && mailtimes.com.au, portnews.com.au, ...\\
        
        Postmedia Network Canada Corp & 58 & lfpress.com, nationalpost.com,\\
        && windsorstar.com, winnipegsun.com,\\ 
        && intelligencer.ca, coldlakesun.com, ...\\

        Philips & 45 & usa.philips.com, philips.com.br,\\
        && philips.com.mx, philips.com.pk,\\
        && philips.cz, philips.ru, philips.pl ...\\
    \bottomrule
    \end{tabular}
    \label{tab:communities}
\end{table}

We manually examined communities and tried to determine the legal entity operating or even owning all websites in each community. 
We report some of the largest of these communities in Table~\ref{tab:communities} along with their size (\ie/ number of websites) and some indicative websites as examples.

\end{document}